\documentclass{aa}
\usepackage{pictexwd,rotating,amssymb,nicefrac}
\usepackage{booktabs,afterpage}
\usepackage[dvips]{color}
\newfont{\tinytiny}{cmr5 at 4pt}
\newfont{\tinytinytiny}{cmr5 at 1pt}

\def\yerrorbar at #1 #2, yerror from #3 to #4 {%
\put {\line(1,0){1}} at #1 #3
\put {\line(1,0){1}} at #1 #4
\plot #1 #3 #1 #4 /
\put {\circle*{0.5}} [Bl] at #1 #2
}
\begin{document}

   \title{QSOs from the variability and proper motion survey \\
         in the M\,3 field}

   \author{H. Meusinger$\,^{1,}$
   \thanks{
Visiting Astronomer, German-Spanish Astronomical Centre, Calar Alto,
operated by the Max-Planck-Institute for Astronomy, Heidelberg, jointly
with the Spanish National Commission for Astronomy},     
          \ R.-D. Scholz$\,^{2,\, \star\star}$,
          \ M. Irwin$\,^{3}$,
	  \ H. Lehmann$\,^{1,\, \star\star}$
 }

   \institute{
   $^{1}$ Th\"uringer Landessternwarte Tautenburg, 07778 Tautenburg, Germany\\
   $^{2}$ Astrophysikalisches Institut Potsdam, An der Sternwarte 16, 14482
   Potsdam, Germany\\
   $^{3}$ Institute of Astronomy, Madingley Road, Cambridge CB3 1HA, UK
             }

 \offprints{H. Meusinger}

   \date{}

   \abstract{We present results of the spectroscopical follow-up
             observations of QSO candidates from a combined variability
	     and proper motion (VPM) survey in a $\sim 10$ square degrees region
	     centered on the globular cluster M\,3. The search is based on a
	     large number of digitised Schmidt plates with a time-baseline
	     of three decades. This paper reviews the candidate selection,
	     the follow-up spectroscopy, and general
	     properties of the resulting QSO sample.
	     In total, 175 QSOs and Sey1s were identified among the
	     objects from the VPM survey, with 114 QSOs and 10 Sey1s
	     up to the pre-estimated 90\% completeness limit of the survey at
	     $B_{\rm lim} \approx 19.7$. The redshift range of the QSOs
	     is $0.4 <z<3$. Among the 80 QSO candidates
	     of highest priority we confirm 75 QSOs/Sey1s and 2 NELGs.
	     We present magnitudes, colours, redshifts, and variability
	     indices for all 181 identified QSOs/Sey1s/NELGs and
	     spectra for the 77 QSOs/Sey1s/NELGs from our spectroscopic
	     follow-up observations. The VPM
	     survey uses selection criteria that are not directly relying
	     on the spectral energy distribution of QSOs. It is
	     therefore remarkable that the properties of the VPM QSOs
	     do not significantly differ from those of samples from
	     colour selection or slitless spectroscopy.
	     In particular, we do not detect a substantial number of
      unusually red QSOs.
	     The total surface density of the brighter QSOs
	     ($17 \le B \le 18.5$)
	     in our search field is found to be by a factor of $\sim 1.8$
      larger than that derived from previous optical surveys.
   \keywords{Galaxies: active --
             Galaxies: Seyfert --
             Galaxies: statistics --
             quasars: general --
             }
   }

\authorrunning{Meusinger et al.}
\titlerunning{VPM QSO survey in the M\,3 field}

   \maketitle

%
%
\section{Introduction}
%

The variability and proper motion (VPM) survey is a QSO search project
that is based on optical long-term variability and non-detectable 
proper motions. Variability of flux densities and  
stationarity of positions are two fundamental properties of quasars,
and therefore well suited as selection criteria of a QSO search 
(e.\,g., Kron \& Chiu \cite{Kro81}; Hawkins \cite{Haw83}; Majewski et al. \cite{Maj91}; 
V\'eron \& Hawkins \cite{Ver95}; Bershady et al. \cite{Ber98}). However,  
due to the special demands on the number and the 
time-baseline of the available observations such attempts must be limited to
comparatively small and confined areas.  
We performed the VPM technique  in two Schmidt fields of $\sim 10$ \,square 
degrees each on the basis of a large number of altogether more than 200 digitised
Tautenburg Schmidt plates in the B band with a time-baseline of three decades
(Meusinger et al. \cite{Meu97}, \cite{Meu02a}).

It is not the primary aim of this project to increase the number of known QSOs
by an insignificant fraction; the problems of detecting substantial numbers of 
QSOs have long been overcome. Over the last decade, among others,
the Durham/AAT survey (Boyle et al. \cite{Boy90}), 
the Large Bright Quasar Survey (Hewett et al. \cite{Hew95}),
the Edinburgh Quasar Survey (Goldschmidt \& Miller \cite{Gol98}), 
and the Hamburg/ESO survey (Wisotzki et al. \cite{Wis00})
have been completed. Presently, the 2dF Quasar Survey (Croom et al.
\cite{Cro01}) and the Sloan Digital Sky Survey (Schneider et al.
\cite{Sch02}) are extremely efficient at identifying very large 
numbers of quasars. The INT Wide Angle 
Survey (Sharp et al. \cite{Sha01}) is expected to detect a statistically 
significant sample of high-redshift quasars. Very deep quasar samples were
obtained in the Lockman hole via the X-ray satellite ROSAT
(Hasinger et al. \cite{Has98}) and in the optical domain with the
Hubble Space Telescope (e.\,g., Conti et al. \cite{Con98}), 
respectively. Further, the VLA FIRST Bright Quasar Survey 
(e.\,g., White et al. \cite{Whi00})
will define a radio-selected QSO sample that is competitive in
size with current optically selected samples. 

Most of the criteria for the selection of QSO 
candidates rely upon differences in the broad-band spectral
energy distribution of QSOs and stars. Despite the large number 
of QSOs now catalogued, the selection effects of the
conventional surveys are not yet fully understood. 
It is therefore important to perform QSO
surveys that are based on different selection methods.
The VPM survey  does not
directly invoke the spectral energy distribution as the primary
selection criterion and provides therefore an
interesting opportunity to evaluate the selection effects of more
conventional optical QSO searches. 
For instance, a serious question concerns the possible  
existence of a substantial population of red QSOs. 
Extinction-reddened QSOs are suggested both from 
the AGN unification model (e.\,g., Antonucci \cite{Ant93}; 
Maiolino \cite{Mai01}) and from the hypothesis of ultra-luminous IR 
galaxies (ULIRGs) as QSOs in the making (Sanders et al. \cite{San88}).
The vast majority of catalogued QSOs have uniform spectral
energy distributions with a blue continuum and broad absorption lines.
Over the last few years,
QSOs with extreme red colours have been detected on the basis of
their X-ray emission (e.\,g., Risaliti et al. \cite{Ris01})
or by radio surveys (Webster et al. \cite{Web95}; Francis et al.
\cite{Fra00}; White et al. \cite{Whi00}; 
Menou et al. \cite{Men01}; Gregg et al. \cite{Gre02}). 
The fraction of unusually red objects 
among the whole QSO population is however unknown. When compared
to other optical surveys, the VPM technique has the advantage
that it can discover such red QSOs as long as (1.) they are
not too faint in the B band and (2.) they are not much less
variable than the conventional QSOs.

The VPM survey was started in the high-galactic latitude field 
around M\,3 (Meusinger et al. \cite{Meu95}). Half of this field 
is covered by the CFHT blue grens survey (e.\,g., Crampton et al.
\cite{Cra90}). The CFHT QSOs could serve as a training set and were
used to define the selection 
thresholds for the VPM survey in such a way that 
a $\sim90$\% completeness is expected up to $B \approx 19.7$. 
The strategy, the observational material,  and the data reduction 
for the M\,3 field were presented in detail in Paper~1 
(Scholz et al. \cite{Sch97}). 
The procedures and results for the second search field, around M\,92, 
are described in a series of papers (Brunzendorf \& Meusinger \cite{Bru01},
\cite{Bru02}; Meusinger \& Brunzendorf \cite{Meu01}, \cite{Meu02b}). 
A brief review  of the whole VPM project is given by Meusinger et al. 
(\cite{Meu02a}).
The present paper presents the QSO sample in the M\,3 field.    
In Section~2, we briefly discuss the candidate selection.
The spectroscopic follow-up observations are
described in Section~3. Section~4 gives an overview of the properties 
of the QSO sample. Conclusions are given in Section~5. As in the 
previous papers, we adopt $H_0 = 50$\,km\,s$^{-1}$\,Mpc$^{-1}$ and $q_0=0$.

%
\section{QSO candidate selection}
%


\begin{figure}[hpbt]
\resizebox{8.8cm}{10.8cm}{\includegraphics{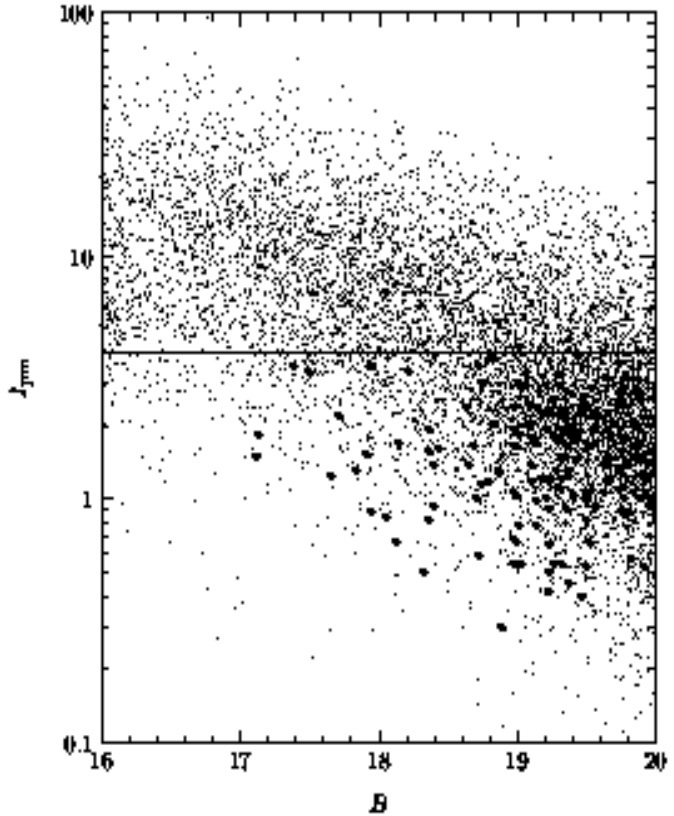}}
\caption{\label{pm_b}
Proper motion index $I_{\rm pm}$ as a function of the mean
$B$ magnitude for about 7\,000 star-like objects with $17 < B < 20$ 
(small grey dots). The 154 QSOs, Sey1s, and NELGs in this magnitude 
range are marked as bullets.
The horizontal line indicates the proper motion selection threshold.
}
\end{figure}

The VPM search is based on indices for
star-like image structure, positional stationarity,
overall variability, and long-term variability
measured on 
57 $B$ plates taken with the Tautenburg Schmidt plates 
between 1964 and 1994, i.\,e. with a time-baseline of three decades.
The $B$ magnitudes given in the present paper are mean magnitudes 
from this database.
The strategy for the QSO candidate selection of the VPM search
in the M~3 field and the definitions of the indices are outlined
at length in Paper~1. Here we review only the basic ideas and
describe the modifications in the selection procedures.

The proper motion index, $I_{\rm pm}$,
is expressed simply by the measured proper motion in units
of the proper motion error. The overall variability index, $I_{\rm var}$,
is assessed by the deviation of the individual magnitudes about the mean
magnitude, and is normalised by the average magnitude scatter
for star-like objects in the same magnitude range. Finally, an
index for long-term variability, $I_{\rm ltvar}$, is defined
by means of structure function analysis and is computed for all 
star-like objects with $I_{\rm pm} < 4$ (see below) and $B<20$.
The selection thresholds for the indices were derived
from the statistics of the previously known QSOs in the field.
There are 90 such QSOs identified with star-like objects measured on
at least 7 of our $B$ plates.
We found that a good compromise between the success rate
(i.e., the fraction of candidates that turn out to be QSOs)
and the completeness (i.e., the fraction of all QSOs found by
the survey) is achieved for the following set of constraints:
$I_{\rm pm}<4,\, I_{\rm var}>1.3$,
and $I_{\rm ltvar}>1.4$. (In Paper~1, the long-term variability
index was denoted $RS_{100}$.) The pre-estimated values
for the success rate and the completeness are 90\% and 40\%,
respectively, for a limiting magnitude $B_{\rm lim} = 19.7$.
(Note that the limiting magnitudes of the individual $B$ plates
vary from 19.5 to 21.3.)

For stationary objects, the probability $p_{\rm pm}$ to measure a
non-zero proper motion follows a Weibull-distribution and
depends only on the proper motion index (Brunzendorf \& Meusinger
2001). An object with $I_{\rm pm} \ge 4$ has a probability
of $p\ge 0.9997$ for non-zero proper motion.
As illustrated by Fig.\,\ref{pm_b}, the proper motion
selection is in particular efficient for brighter magnitudes
where the proper motion errors are smaller. For $B < 18.5$,
the typical proper motion error is about 1\,mas\,yr$^{-1}$, and
83\% of the star-like objects have $I_{\rm pm} > 4$.
For $19<B<20$, on the other hand, the typical proper motion
error is about 3\,mas\,yr$^{-1}$, and only about 21\% of the star-like
objects have $I_{\rm pm} > 4$.
In a flux-limited sample, most of the objects have magnitudes
close by the limit. Hence, the proper motion selection appears
not very efficient for the whole survey with a limit at $B \approx 19.7$.
However, at brighter magnitudes, the number-magnitude relation
for QSOs is much steeper than for the foreground stars.
This means that the contamination of the variability-selected
QSO candidate sample by foreground stars is stronger at
brighter magnitudes where the proper motion selection works
more efficiently. At fainter magnitudes,
the zero proper motion constraint is important in particular
for the efficient rejection of nearby variable late-type
main sequence stars (see Paper~1).

The selection starts with 32\,700 objects detected on a deep master
plate. About 24\,600 objects were identified on at least two further plates,
among them are about 21\,500 objects with star-like images.
A basic object sample for the variability selection is defined
by the 12\,800 star-like objects measured on at least 7 $B$ plates.
After excluding
the objects in the crowded cluster region (distance to the centre
of M\,3 less than $24'$), this sample is reduced to 8\,582 objects
in total and to 4\,614 objects in the magnitude range $16.5\le B \le 19.7$,
respectively. About 65\% of the objects from this reduced sample
are rejected due to the zero proper motion constraint $I_{\rm pm} < 4$.
Finally, the variability constraints strongly reduce the candidate
sample to a manageable size.

\begin{table}
\begin{tabular}{llll}
\toprule
priority class        & high    & medium    & low\\
\midrule
number of candidates  & 80      & 95        & 607\\
already catalogued    & 26      & 17        &  15\\
newly observed        & 54      & 68        &  27\\
\midrule
QSOs/Sey1s            & 75      & 36        &  20\\
NELGs                 &  2      &  -        &   1\\
stars                 &  3      & 49        &  21\\
\bottomrule
\end{tabular}
\caption{\label{selection} Object numbers for the subsamples of QSO
candidates with high, medium, or low priority.
}
\end{table}

\begin{figure*}[hpbt]
\resizebox{17.8cm}{17.8cm}{\includegraphics{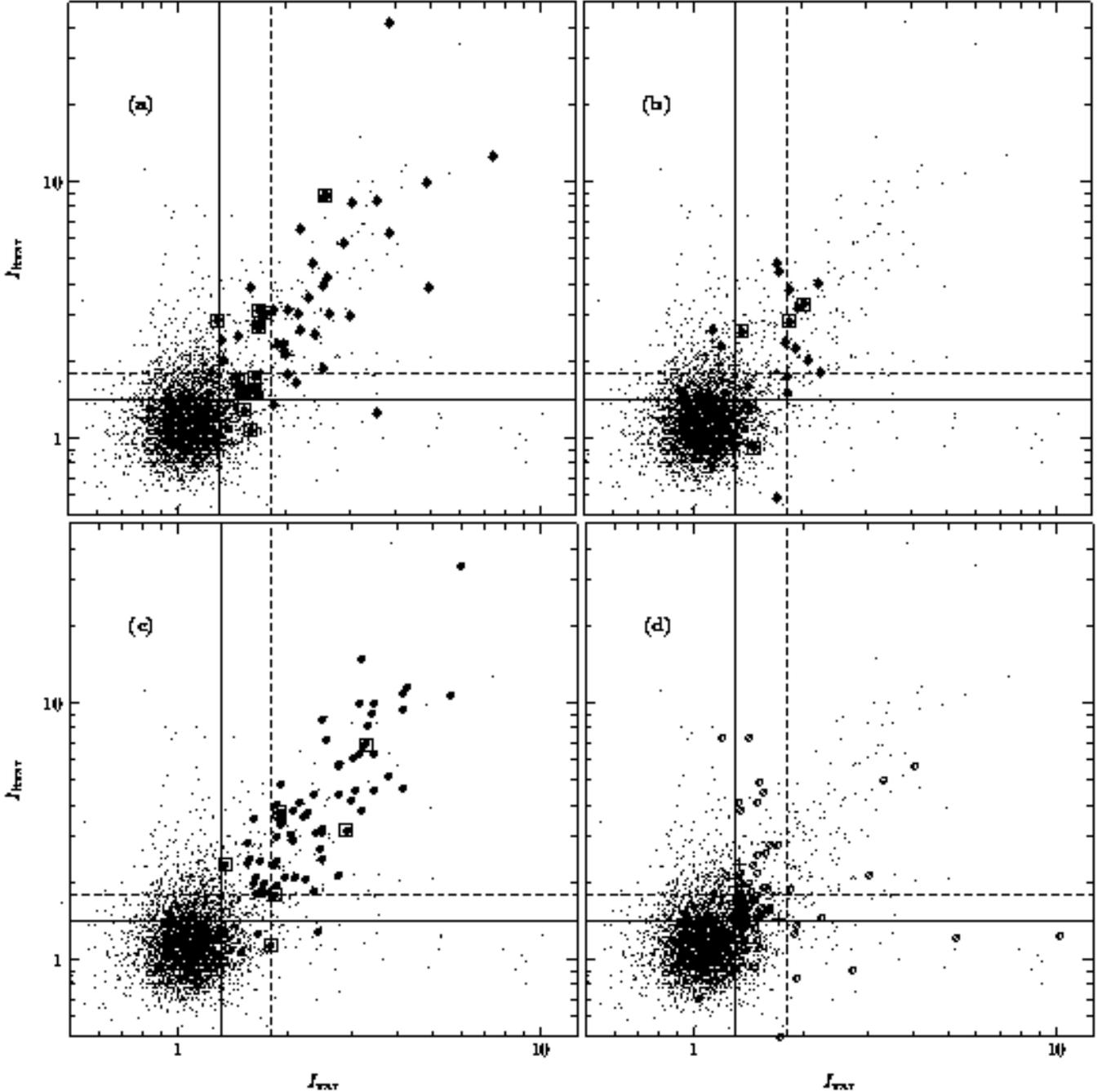}}
\caption{\label{ltv_vi}
Long term variability index $I_{\rm lt var}$ versus overall variability 
index $I_{\rm var}$ for the 4\,379 star-like objects from the
reduced basic object sample with $14 < B < 20$ (small grey dots).  
The previously known QSOs, Sey1s, and NELGs in the field 
are shown as $\blacklozenge$ in the first two panels ({\bf (a)} 
for $B\le 19.7$ and {\bf (b)} for $19.7 < B \le 20$).  
In panel {\bf (c)}, the objects from Table\,3 
are indicated by $\bullet$. 
The symbols for QSOs with $z>2.2$ are framed.
Panel {\bf (d)} shows the QSO candidates from the present study that were 
proved to be 
foreground stars ($\circ$) as well as the 10 medium-priority candidates 
without follow-up spectra ($+$). The lines indicate the variability 
selection criteria (see text). 
}
\end{figure*}

For practical reasons, the candidate sample is devided into
three subsamples of different priority. A similar approach was
used for the VPM search in the M\,92 field
(cf. Brunzendorf \& Meusinger 2001). However, the
variability indices defined there are slightly different from
those used in the present study, and the priority classification
in the two VPM fields are not completely identical.
Here, the priority depends mainly on the variability indices.
In addition, the $B$ magnitudes and the crowding of the
field (expressed by the distance $d_{\rm c}$ from the centre
of M~3) are taken into account. For all priority classes,
star-like objects are considered with $I_{\rm pm} < 4,\, B=15-19.7$,
and $d_{\rm c}>24$\,arcmin. The {\em high-priority subsample}
consists of the strongly variable objects
with $I_{\rm var} > 1.8,\, I_{\rm ltvar} > 1.8$.
The {\em medium-priority subsample} contains the objects with 
smaller variability indices $I_{\rm var} = 1.3-1.8$ and
$I_{\rm ltvar} > 1.4-1.8$. In addition, we included the few objects 
with somewhat higher variabiliy indices ($I_{\rm var} = 1.6-1.8$ and
$I_{\rm ltvar} = 1.6-1.8$) in the stronger crowded region
$d_{\rm c}= 12 - 24$\,arcmin.
Finally, the {\em low-priority subsample} comprises
the objects having only one
of the two variability indices above the threshold
(i.e., $I_{\rm var} > 1.3$ or $I_{\rm ltvar} > 1.4$).
In addition, we consider also objects with
$19.7\le B\le19.8$ and $d_{\rm c} > 12$\,arcmin as low-priority
candidates if at least one of the two variability indices
exceeds the threshold.
The variability selection is illustrated by Fig.\,\ref{ltv_vi}.

As discussed in Paper~1, the U band variability index may serve as 
an additional selection criterion. In practice, however, the 
fainter objects are measured on only a small number of U plates.
Therefore, the U variability index was invoked only 
in one case: the QSO No.\,51 from Table\,3  
has insignificant B variability indices but shows significant 
variability in the U band.

The numbers of selected candidates are 80,
95, and 607, respectively, for the subsamples of high,
medium, or low priority (Table\,\ref{selection}). 
It is
ecpected that the fraction of QSOs/Sey1s strongly decreases
with decreasing priority. In particular, the low-priority
subsample is ecpected to be strongly contaminated by 
galactic stars with relatively enhanced photometric errors. 

A cross-check of the candidate list against the
NED\footnote{The NASA/IPAC extragalactic database (NED)
is operated by the Jet Propulsion Laboratory, California
Institute of Technology, under contract with the National
Aeronautics and Space Administration.} (2002, February)
yields the identification (identification radius 10\,arcsec)
of 57 QSOs/Sey1s and one narrow emission line galaxie (NELG)
with catalogued redshifts.
Over the whole magnitude range, 104 objects with catalogued
redshifts $z>0$ were identified (100 QSOs/Sey1s, 4 NELGs).
The overwhelming fraction of the QSOs/Sey1s are from the CFHT
blue grens survey (e.\,g., Crampton et al. \cite{Cra90})
which covers approximately half of our survey field.

%
\section{Spectrosopic follow-up observations}
%

Spectroscopic follow-up observations have been focussed on the
candidates of high or medium priority. Most of the spectra were
taken during five observing runs either with the 2.2\,m
telescope on Calar Alto or with the Tautenburg 2\,m telescope.
An overview of these 
 observation runs is given in Table\,\ref{obs_log}.
In addition, three candidates with uncertain spectroscopic identification
were re-observed in July 2001 with CAFOS on Calar Alto; this run is
quoted as number 6 in Table\,3.  
An additional 18 candidates of medium or low priority
were proved to be foreground stars during several other campaigns 
with either the Tautenburg telescope or the Calar Alto 2.2\,m
telescope.

\begin{table*}
\begin{tabular}{lccccc}
\toprule
observing run  &     1      &     2      &     3      &     4      &     5      \\
\midrule
spectrograph     & TAUMOK     & CAFOS      & CAFOS      & CAFOS      & CAFOS \\
year/month       & 1997/04    & 1998/04    & 1999/04    & 2000/04    & 2001/03 \\
number of nights &     7      &     7      &     5      &     3      &     3 \\
number of objects &  41       &    46      &    34      &    23      &   35  \\
\bottomrule
\end{tabular}
\caption{\label{obs_log}
Observation log for the major spectroscopic follow-up observation
runs. 
}
\end{table*}

%
\subsection{\label{taumok}Multi-object spectroscopy with TAUMOK}
%

The brighter candidates ($B<18$) were observed with TAUMOK in the
Schmidt focus of the Tautenburg 2\,m telescope.
TAUMOK allows to obtain simultaneously
spectra of up to 35 objects
within an area of $2\fdg3$ diameter (see Meusinger \& Brunzendorf 
\cite{Meu01} for more details). 
The telescope was operated in a scanning mode prior to the 
spectroscopic observations in order to determine the most accurate
positions of the fibres. 
The wavelength coverage is approximately 3\,800--9\,000\,{\AA}, the
reciprocal linear dispersion is 400\,{\AA}\,mm$^{-1}$ corresponding 
to 9 {\AA} per pixel.

The atmospheric conditions during the TAUMOK campaign were moderate.
In five of the seven nights, spectra of QSO candidates were taken.
Four different fibre configurations were necessary to cover the 
VPM field.  Several 1\,800\,s exposures were taken for each configuration.
The total exposure time per field is between 1.5 and 3\,hours.
Since the number of bright high-priority
candidates is much smaller than the total number of available
object fibres, most of the fibres were positioned at candidates
of lower priority or on non-priority objects. Five fibres were reserved for
template spectra from known QSOs with redshifts beteween $z=0.6$
and 2.5. Internal spectral lamps were used for the wavelength calibration 
prior and after the observation of a field.
For the reduction of the TAUMOK spectra we applied a 
software package (Ball \cite{Bal00}) which is based on 
IRAF standard procedures for multi-object spectroscopy.

%
\subsection{\label{cafos}Single-object spectroscopy with CAFOS}
%

The candidates fainter than $B\approx18$ were observed with the
focal reducer and faint object spectrograph CAFOS at the
Cassegrain-focus of the 2.2\,m telescope of the German-Spanish
Astronomical Centre on Calar Alto, Spain. The B\,400 grism was
used  with a wavelength coverage of about 3\,000--9\,000\,{\AA}.
The width of the entrance slit was adjusted to the seeing
(typically 2-3\,arcsec) resulting in an effective linear
resolution of typically about 40-60\,{\AA}. Since the orientation of the
long-slit was always North-South, some ``slit-loss'' due to 
atmospheric dispersion was unavoidable 
for spectra taken at hour angles significantly different from zero.

The total integration time varied between 10 and 90\,min,
dependent primarily on the strength of the emission lines and
on the weather conditions. Observations were made in a wide variety
of atmospheric conditions. The weather was good in the 2001 observing
campaign. In the previous runs, however,
the fraction of observing time with good atmospheric transparency
was rather low. Therefore, about half of all spectra have only a moderate 
signal-to-noise ratio. Several objects had to be observed
in more than one run.
Data reduction was performed using the long-slit
spectroscopy package {\tt LONG} of MIDAS. Wavelength calibration was 
done by means of calibration lamp spectra. 

%
\subsection{\label{overview}Overview of the spectroscopically observed objects}
%


\begin{figure*}[hpbt]
\resizebox{18.0cm}{9.5cm}{\includegraphics{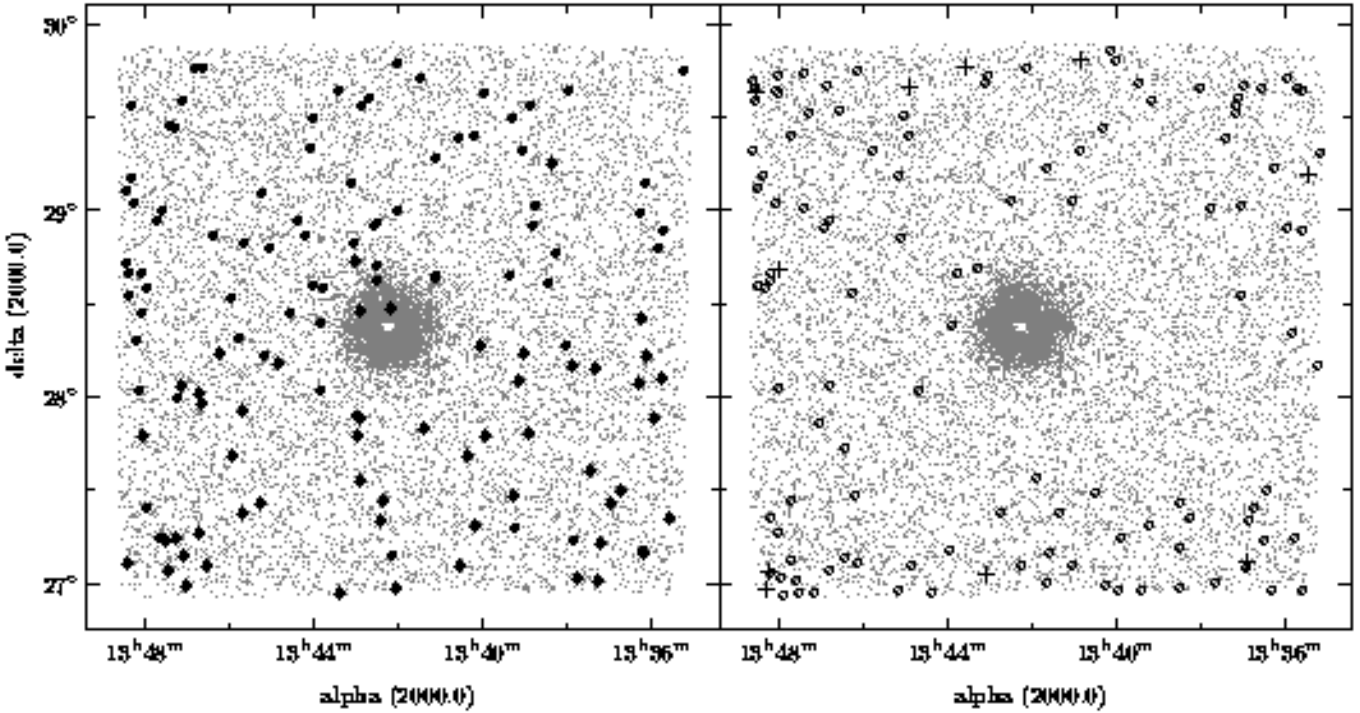}}
\caption{\label{map}
Field of the VPM survey centred on the globular cluster M\,3. 
Star-like objects are shown as grey dots. The panel on the left hand side
shows the distribution of the QSOs from the VPM search ($\bullet$) 
as well as the QSOs with $B<19.7$ from the NED ({\tiny$\blacklozenge$}).
The panel on the right hand side shows  the distribution of the VPM 
candidates that were spectroscopically confirmed to be foreground stars 
($\circ$) as well as the medium-priority candidates without follow-up 
spectra ($+$). 
}
\end{figure*}

Spectra were taken for a total of 198 objects:
{\it 1.)} In particular, all 54
high-priority candidates without classification in the NED were observed.
{\it 2.)} Among the 95 medium-priority candidates, 17 objects have a measured
redshift in the NED. For an additional 68 medium-priority objects
spectra were taken in the framework of the present study. 
The remaining 10 objects have variabilities near
the selection thresholds (Fig.\,\ref{ltv_vi}d)
and/or are located at the borders of the field (Fig.\,\ref{map})
where the contamination by foreground stars is obviously stronger
(see below).  The chance to find QSOs among these remaining 10 
candidates is substantially lower than for the 
median-priority subsample as a whole.
{\it 3.)} The number of low-priority candidates is obviously too large
to reach a substantial completeness with regard to spectroscopic follow-up
observations.
We selected therefore from this priority class mainly the brighter objects
($18< B <19.5$) and/or the objects with the strongest indication for
variability.
Note that the high fraction of QSOs/Sey1s among the {\it observed} 
 low-priority objects is thus not representative for the whole 
 subsample. There may be some undetected QSOs among the remaining 
 low-priority objects, but their number is expected to be small.

The statistics of the observations for the various priority classes
are summarised in Table\,\ref{selection}. Note that the total
number of observed candidates listed there is smaller than the
number of all observed objects given in Table\,\ref{obs_log}. The
reasons for this apparent discrepancy are the following:
{\it (a)} the criteria for the definition of priorities have
slightly changed during the survey.
{\it (b)} Many of the brighter objects observed with TAUMOK are not candidates
in the sense of Table\,\ref{selection}, but were selected to allow
a good positioning of the fibres.
{\it (c)} Several objects were observed in more than one 
 observing run.
{\it (d)} Since a variability survey is expected to be biased against
low-variability QSOs, we selected also a few objects with quasar-typical
colours, but with variability indices slightly below the selection
thresholds. For instance, four objects were observed because they are X-ray 
sources. {\it (e)} A few strongly variable objects with $19.7<B<19.9$ 
have been observed as well.

%
\subsection{Source classification}
%

The spectral classification is based on the emission and
absorption line properties. Three catagories are considered:
(1.) redshifted broad emission lines and/or absorption lines, 
(2.) redshifted narrow emission lines, and
(3.) unredshifted typical stellar absorption lines.

The first category comprises QSOs and Sey1s, which are
discriminated by the usual luminosity threshold $M_{\rm B} = -23$.
The absolute magnitude $M_{\rm B}$ is computed for 
$H_0 = 50$\,km\,s$^{-1}$\,Mpc$^{-1},\, q_0 = 0$ and the
$K_{\rm B}$-correction from Brunzendorf \& Meusinger (\cite{Bru01}).
The data for the 69 QSOs and 5 Sey1s from the present study are listed in
Table\,3,  
and in the following, objects will be quoted with
their number in this table. For most of these objects, redshifts
were derived from several emission lines; in particular, strong
narrow forbidden lines
(e. g., [\ion{O}{iii}]$\lambda5007$) were used, if present.
The wavelength of a single line was measured by Gaussian centroids.

When we selected the first QSO candidate list in 1996, this list has
been checked against the NED (cf. Paper\,1) to reject all objects
already catalogued with measured redshifts. A new check revealed that six
QSOs from Table\,3  
are catalogued in the February 2002 release
of the NED. For three of these objects (No. 16, 68, 70) redshifts were
published  later than 1996. (Number\,68 is
identical with FBQS\,J1348+2840, White et al. \cite{Whi00}.)
The other three (No. 29, 36, 42) had uncertain
positions in the 1996 NED and therewith too large position differences
($> 10$\,arcsec) for an unambiguous identification. Further, we
 note that the QSO No.\,11 is identified with the radio source
(without $z$ in the NED) FIRST\,J133825.6+283637.

There are four narrow-emission line galaxies (NELGs)
among the identified objects from the NED. An additional three NELGs were
detected in the present study and are also listed in
Table\,4. 
The luminosities of the NELGs are
clearly below the QSO-Sey1 threshold. In general,
the class of the NELGs  includes Seyfert\,2s, narrow-line Seyfert\,1s,
LINERs, and \ion{H}{ii} galaxies.
For this paper we have not attempted to separate the types of NELGs.
One of the new NELGs (No. 43) is a high-priority QSO
candidate, another one (No. 26) is of low priority, but with a high overall
variability index $I_{\rm var} = 2.45$. The third one (No.15) has a high
proper motion index and is not a QSO candidate from the VPM survey, but
is one of the X-ray sources observed for completing the QSO sample.
Among all objects with $z>0$  from our basic sample, No.15
is the only one with a proper motion index significantly larger
than the selection threshold $I_{\rm pm}=4$ (Fig.\,\ref{pm_b}),
perhaps indicating a wrong spectral classification from a noisy spectrum.
All 7 NELGs were classified as star-like objects
on the Schmidt plates; the infered redshifts are between 0.137 and 0.433.
In the frame of the VPM search in the M\,92 field, a higher fraction of
NELGs was detected due to  a less stringent star-galaxy separation
(Meusinger \& Brunzendorf \cite{Meu01}). The high variability indices
measured for the NELGs were explained by increased photometric errors for
objects with image profiles deviating from stellar ones (Meusinger \&
Brunzendorf \cite{Meu02b}).

Finally, an object is classified as a foreground star if its spectrum 
unambiguously shows typical un-redshifted stellar absorption lines.
At a first glance, most of these objects are normal stars
without unusual spectral features. 
 Contrary to the QSOs, the classified
stars show a remarkably inhomogeneous distribution over the field
(Fig.\,\ref{map}): their strong concentration towards the
outer parts and the corners of the field indicates an increase
of the instrumental variability at large distances from the plate centre.
Such an effect is in principle expected since we have not corrected
for a position-dependence of the magnitude scale (Paper\,1).
This interpretation implies that a substantial fraction of the
selected stars are not really variables. In this context
we note that most of the stars have lower variability indices than
the QSOs (Fig.\,\ref{ltv_vi}d).

To summarise, we have plausibly classified all 198 objects from our
spectroscopic follow-up observations as either QSOs/Sey1s, NELGs, or
foreground stars. There are new redshifts for 68
broad-lined objects and 3 narrow-lined objects. For an additional
6 already catalogued QSOs/Sey1s redshifts were confirmed.

%
\section{\label{sample}Properties of the QSO sample}
%

%
\subsection{\label{general}General}
%

Table\,3 
lists redshifts, absolute magnitudes, colours,
proper motion indices, and the two variability indices of  
the 77 QSOs, Sey1s, and NELGs from our follow-up spectroscopy.
In a similar style, Table\,4  
summarises the data for
the 104 QSOs, Sey1s, and NELGs identified in the NED. The distribution 
of these types over the three priority classes from the VPM survey 
is given in  Table\,4. 
In the high priority subsample, 94\% of the candidates were found
to be QSOs/Sey1s, while the contamination by foreground stars is
as low as 4\%. For the combined sample of high-and-medium-priority
objects the success rate (i.\,e., the fraction of established QSOs/Sey1s among
all candidates) is still as high as 63\%.

\begin{figure}[hpbt]
\resizebox{8.8cm}{6.8cm}{\includegraphics{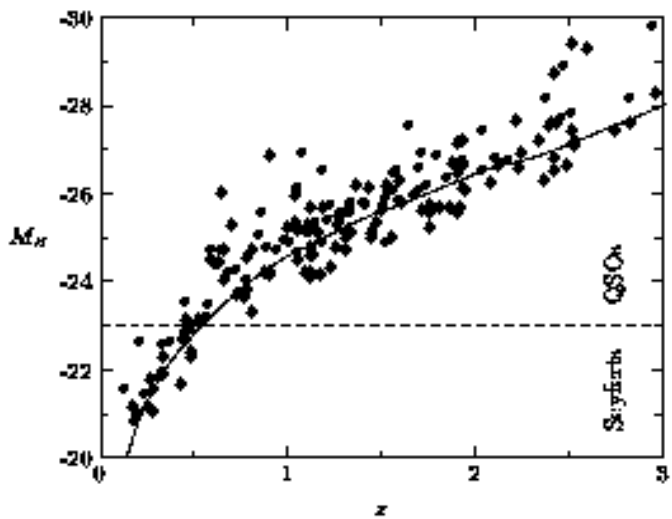}}
\caption{\label{mb_z}
Absolute magnitude $M_{\rm B}$ versus redshift $z$
for all known QSOs/Sey1s in the field identified with objects
from our basic sample. The redshifts are either from
the present study ($\bullet$) or the NED ({\tiny$\blacklozenge$}).
The continuous line indicates the magnitude 
limit $B\le19.7$ of the survey.
}
\end{figure}

Figure\,\ref{ltv_vi} illustrates that a high fraction of all QSOs
in the field are strongly variable.  The $B$ standard deviation due to
variability is about 0.2\,mag for QSOs with $B<19.7$. In this magnitude 
range, more than 60\% of the QSOs/Sey1s show the strong variability of 
high-priority VPM candidates. (A detailed analysis of the
variability properties will be deferred to a separate study.)
For 90\% of QSOs/Sey1s, both variability indices are above the
selection thresholds. We find that 50 out of the 53 NED 
QSOs/Sey1s with $B<19.7$ match the selection criteria of our survey, 
corresponding to a completeness of 94\% for the VPM survey. 
Only for two objects both variability indices fall below the selection 
thresholds; another one has a proper motion index slightly above the
threshold. 
The subsample of the 114 QSOs with $B<19.7$ is considered nearly complete.
These QSOs are homogeneously distributed over the search field 
(Fig.\,\ref{map}). In particular, the
QSO surface density in the northern half of the field, which
is not covered by the CFHT blue grens survey, is comparable to that in the
southern part where almost all known QSOs are in the CFHT survey.
An additional three QSOs were detected by the VPM search
in the CFHT field. Two of them (No. 5,15) have ``normal'' spectra and
were obviously ignored by chance in the CFHT survey; the other one (No.\,7)
shows very strong broad absorption line (BAL) features.
The subsample is of course flux-limited, and $M_{\rm B}$ is therefore
strongly correlated with $z$ (Fig.\,\ref{mb_z}). Only for
$z\le0.55$, the subsample is complete with regard to luminosities
(since $B<19.7$ for QSOs of such $z$). Note that most of the
objects with $z<0.55$ shown in Fig.\,\ref{mb_z} are Sey1s.

\begin{figure*}[hpbt]
\resizebox{18.0cm}{6.8cm}{\includegraphics{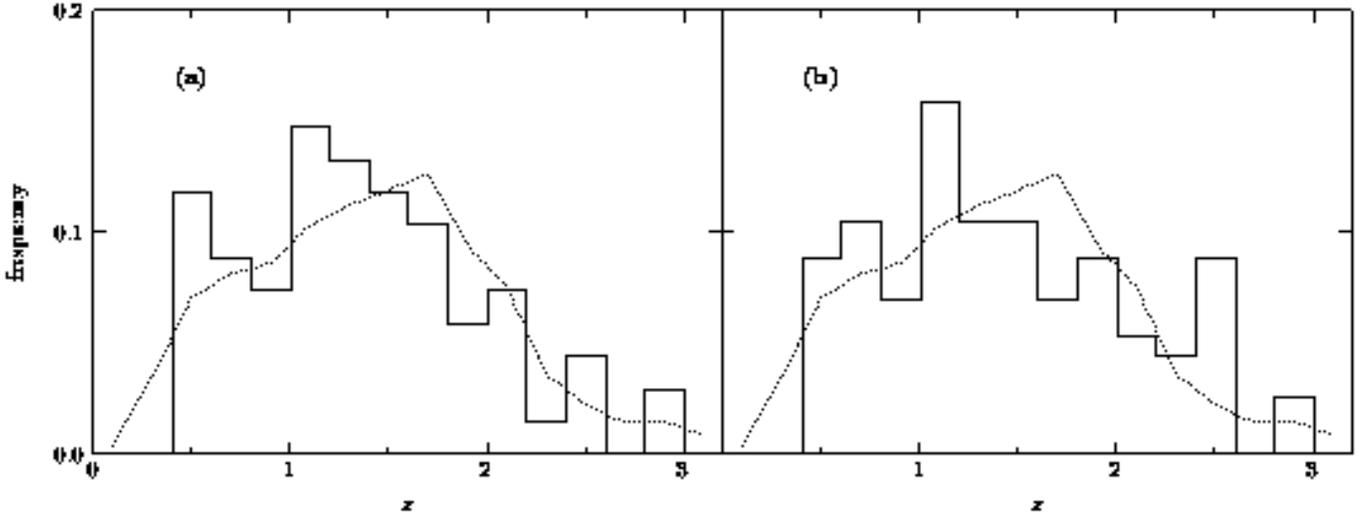}}
\caption{\label{z_hist}
Normalised distribution of the redshifts (number of QSOs per 0.2 
redshift bin) for
{\bf (a)} the subsample from the present follow-up spectroscopy
(Table\,3), and 
{\bf (b)} the (nearly) complete subsample of all 114 QSOs with $B\le19.7$.
For comparison, the dotted polygon gives the normalized $z$ distribution 
for the QSOs from the  SDSS Early Data Release
(Schneider et al. \cite{Sch02}).
}
\end{figure*}

%
\subsection{\label{redshifts}Redshift distribution}
%

The redshift distribution is shown in Fig.\,\ref{z_hist}
both for (a) the subsample from Table\,3  
and (b) the sample of
all identified QSOs with $B\le19.7$. The shape of
the $z$ distribution is roughly comparable with that from the SDSS Quasar
Catalogue I. Early Data Release (Schneider et al. \cite{Sch02}),
corroborating the result from VPM survey in the M\,92 field
(Brunzendorf \& Meusinger \cite{Bru02}). 
This impression is confirmed by the two-tailed KS two-sample test. 
According to this test on a significance level $\alpha=0.05$, we have 
not to reject the null hypothesis that our subsamples (a) and (b) 
and the SDSS sample were drawn from the same population.

%
\subsection{\label{surf_dens}Surface density}
%

\begin{figure}[hpbt]
\resizebox{8.8cm}{6.5cm}{\includegraphics{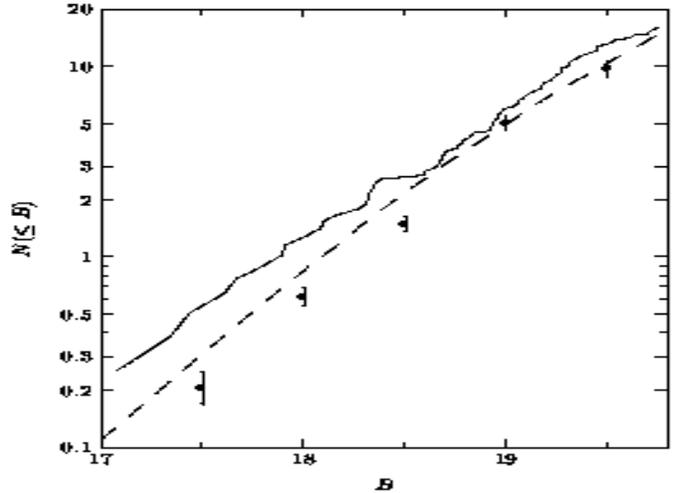}}
\caption{\label{fl_dichte}
Cumulative QSO surface density $N (\le B)$, i.\,e. number of QSOs 
brighter than a given magnitude $B$ per square degree where
$B$ has been corrected for an interstellar
extinction of $A_B=0.05\,$mag. 
Solid polygon: all QSOs outside the cluster region ($d_{\rm c} > 24'$).
Bullets with error bars: integral surface densities from 
Hartwick \& Schade (\cite{Har90}).
The long-dashed curve corresponds to an analytical approximation
given by Wisotzki (\cite{Wis98}) that was derived from a composite 
optical QSO sample.
}
\end{figure}

The ``completeness'', or absolute efficiency, of the survey can
be estimated by comparing the QSO surface densities, i.e. number counts
per solid angle, to the densities predicted by other surveys.
Figure\,\ref{fl_dichte} shows the surface density of all QSOs
(i.\,e., $z>0,\ M_{\rm B}<-23$) with $B<19.7$ in our search field,
compared with mean relations  from
various data samples. The cumulative density $N(<B)$ is simply computed by
dividing the number of QSOs brighter than a given magnitude by the
effective search area where the $B$ magnitudes were corrected for an
 interstellar extinction of $A_{\rm B} = 0.05$\,mag. 
The size of the
Schmidt field is $3\fdg3\times3\fdg$3. After subtracting the
areas of the plate margins
(not shown in Fig.\,\ref{map}), the calibration wedge, the crowded inner
part of M\,3, and the area covered by the images of the objects
in the remaining field, the effective search area amounts to 7.8\,deg$^2$.

The resulting number-magnitude relation is roughly described
by $\log\,N(<B) \propto xB$ with $x\approx0.6$ for $17.5<B<18.5$
and $x\approx0.75$ for $18.5<B<19.5$, in agreement with
the result from the M\,92 field
(Brunzendorf \& Meusinger \cite{Bru02}). The surface densities
for our total QSO sample are higher than those
derived by Hartwick \& Schade (\cite{Har90}), especially at brighter
magnitudes. There are 9 QSOs with $B<18$ in our search field,
corresponding to 1.15\,QSOs\,deg$^{-2}$\,mag$^{-1}$, i.\,e.
a factor of 1.8 more than in the Hartwick \& Schade data.
More recently, La\,Franca \& Cristiani (\cite{LaF97})
derived surface densities of 0.76\,QSOs\,deg$^{-2}$\,0.5\,mag$^{-1}$
for $17.9<B<18.4,\ 0.3<z<2.2,$ and $M_{\rm B}<-23$, to be compared with
1.28\,QSOs\,deg$^{-2}$\,0.5\,mag$^{-1}$ for our sample.
The surface densities based on single-epoch observations are affected
by variability and cannot be compared directly to those
based on time-averaged magnitudes. It should be noticed however
that Hartwick \& Schade corrected their data for such a
variability-induced over-completeness. Note also that
the different assumptions for $q_0$ (both Hartwick \& Schade and
La\,Franca \& Cristiani adopted $q_0=0.5$ while we used $q_0=0$)
make no significant difference for the number counts.
We cannot exclude that the relative overabundance of apparently
bright QSOs is due to the limitations of small-number
statistics, but note that a similar result was
found for the VPM survey in the M\,92 field
(Meusinger \& Brunzendorf \cite{Meu01}).


\begin{figure*}[hbpt]
\includegraphics{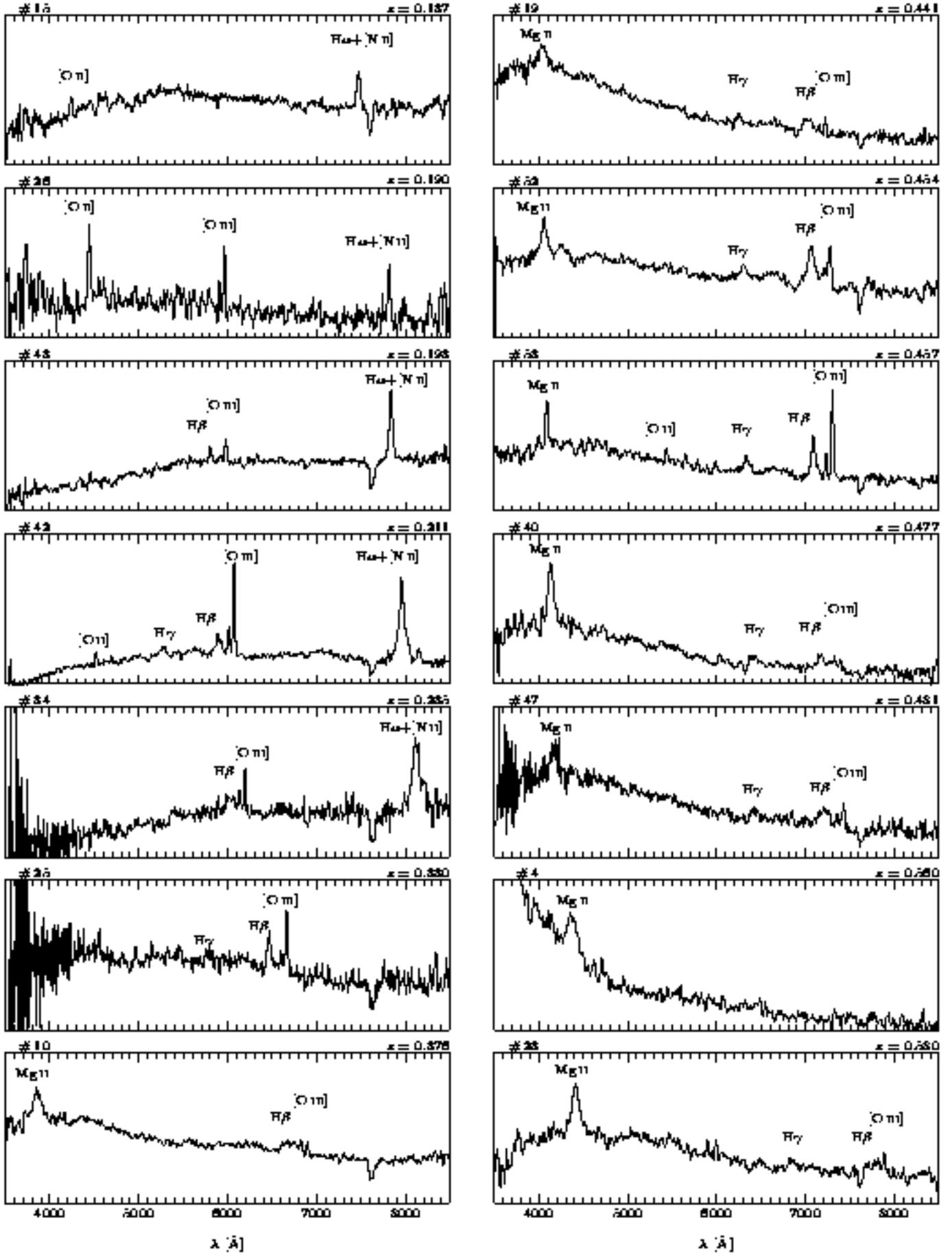}
\caption{\label{allspectra}
Spectra (normalized flux $f_{\lambda}$ versus wavelength $\lambda$) 
of the QSOs, Sey1s, and NELGs 
from Table\,3, sorted by increasing redshift. The running number 
from Table\,3 and the redshift are given for each spectrum. 
Note that the spectra were not corrected 
for the atmospheric absorption bands at 6880 and 7620\,{\AA}. }
\end{figure*}
\afterpage
 
\begin{figure*}[hbpt]
\includegraphics{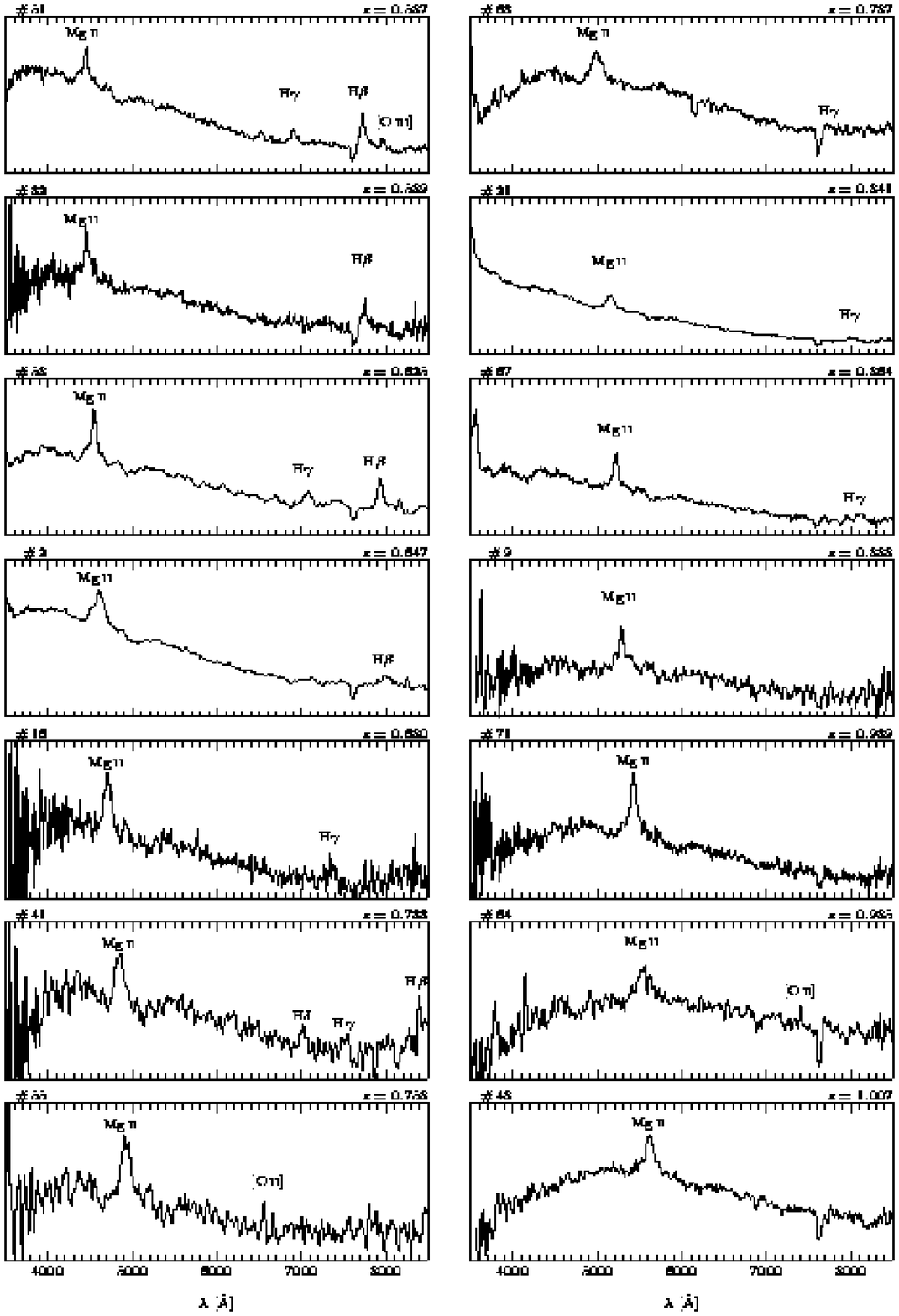}
\addtocounter{figure}{-1}
\caption{(continued)}
\end{figure*}
\afterpage

\begin{figure*}[hbpt]
\includegraphics{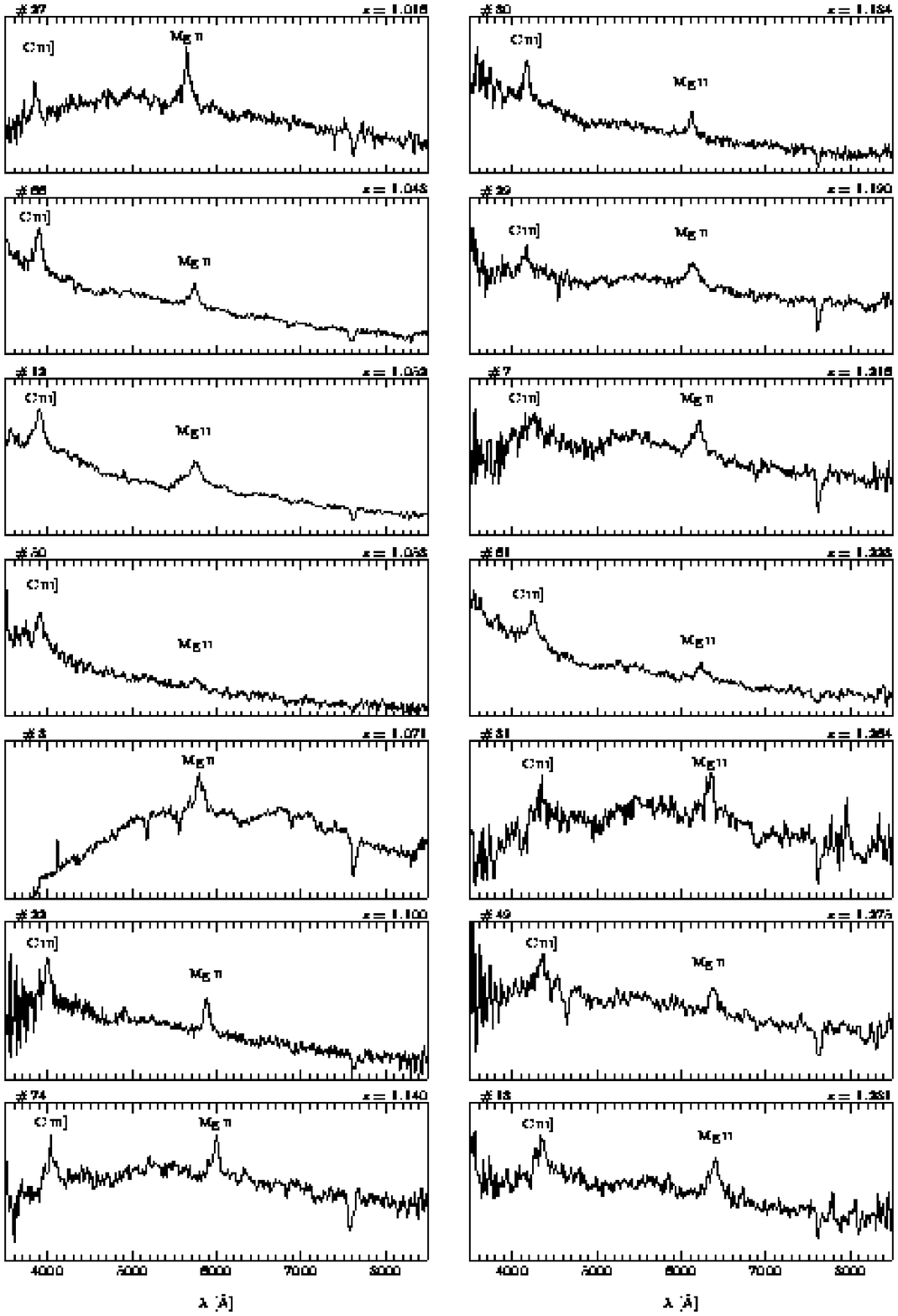}
\addtocounter{figure}{-1}
\caption{(continued)}
\end{figure*}
\afterpage
 
\begin{figure*}[hbpt]
\includegraphics{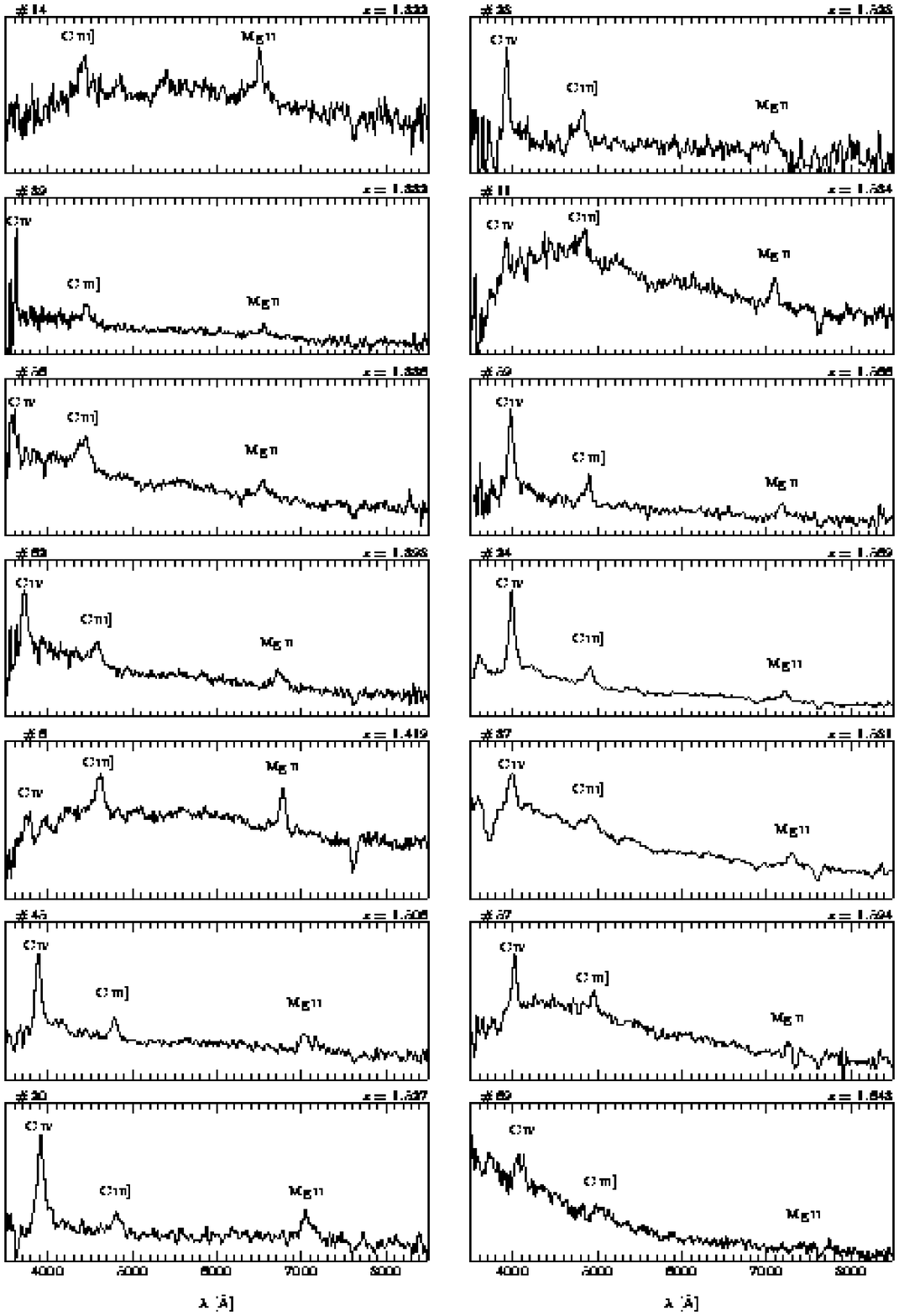}
\addtocounter{figure}{-1}
\caption{(continued)}
\end{figure*}
\afterpage 
 
\begin{figure*}[hbpt]
\includegraphics{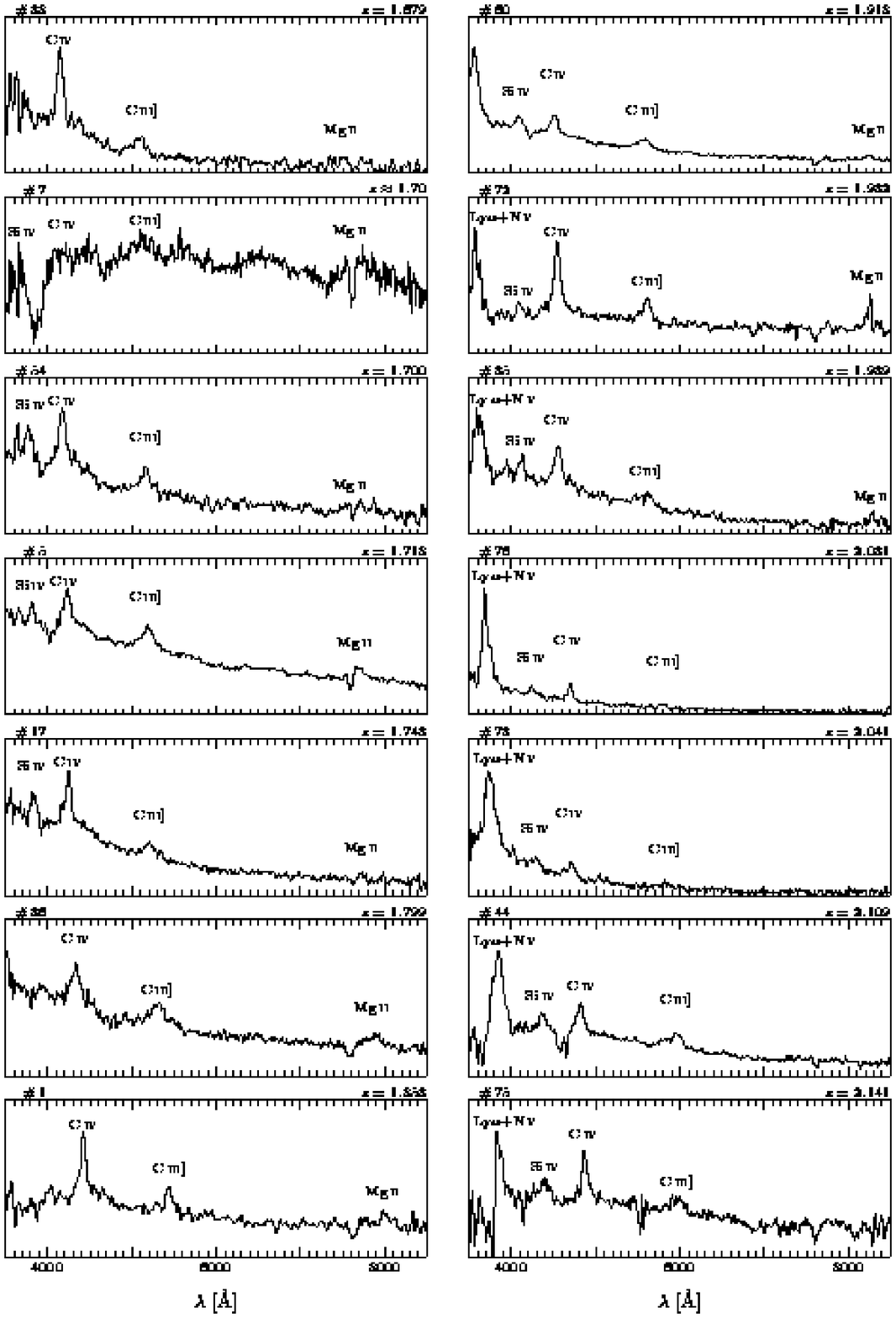}
\addtocounter{figure}{-1}
\caption{(continued)}
\end{figure*}
\afterpage

\begin{figure*}[hbpt]
\includegraphics{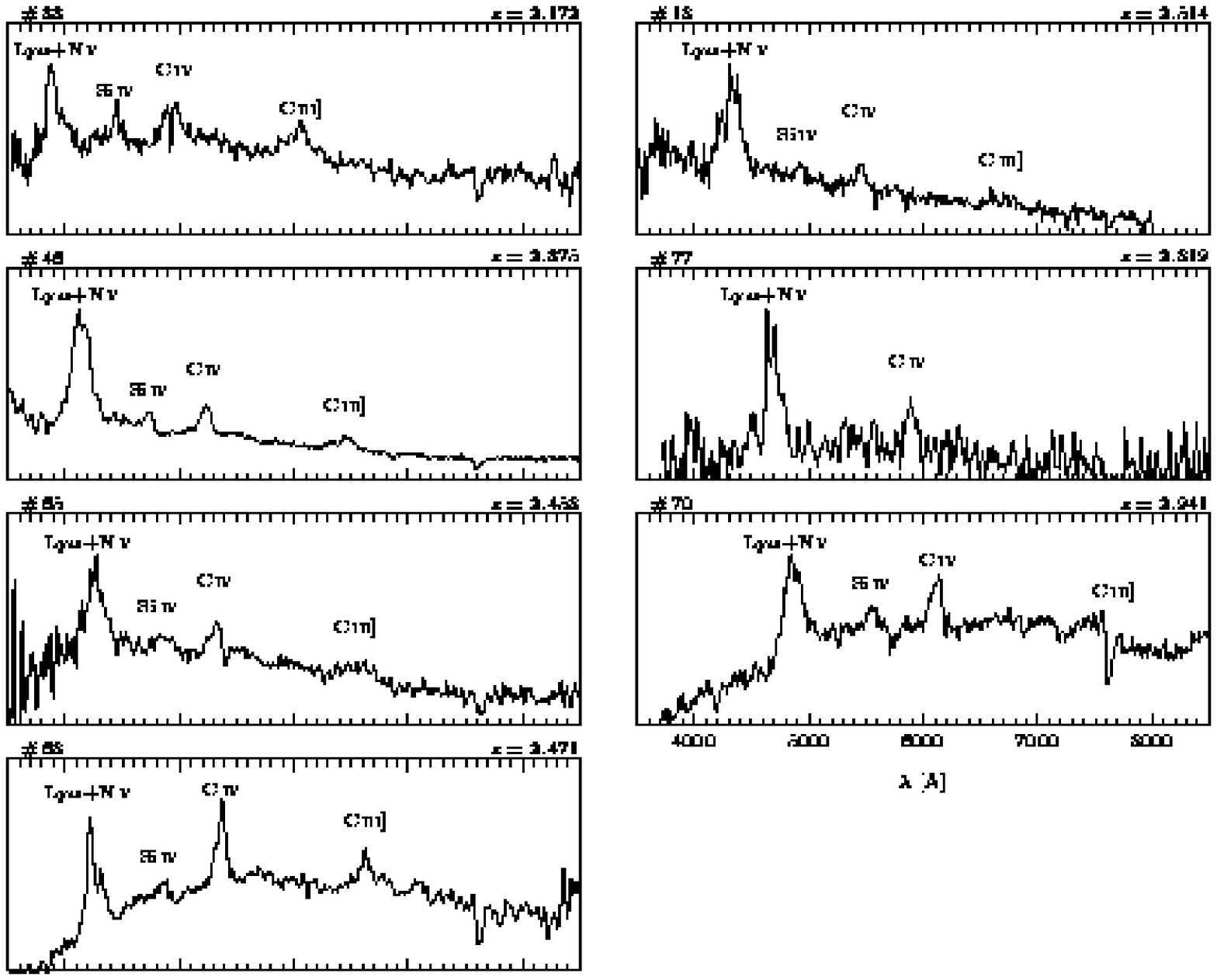}
\addtocounter{figure}{-1}
\caption{(continued)}
\end{figure*}
\afterpage

\begin{figure*}[ht]
\resizebox{16.5cm}{18.8cm}{\includegraphics{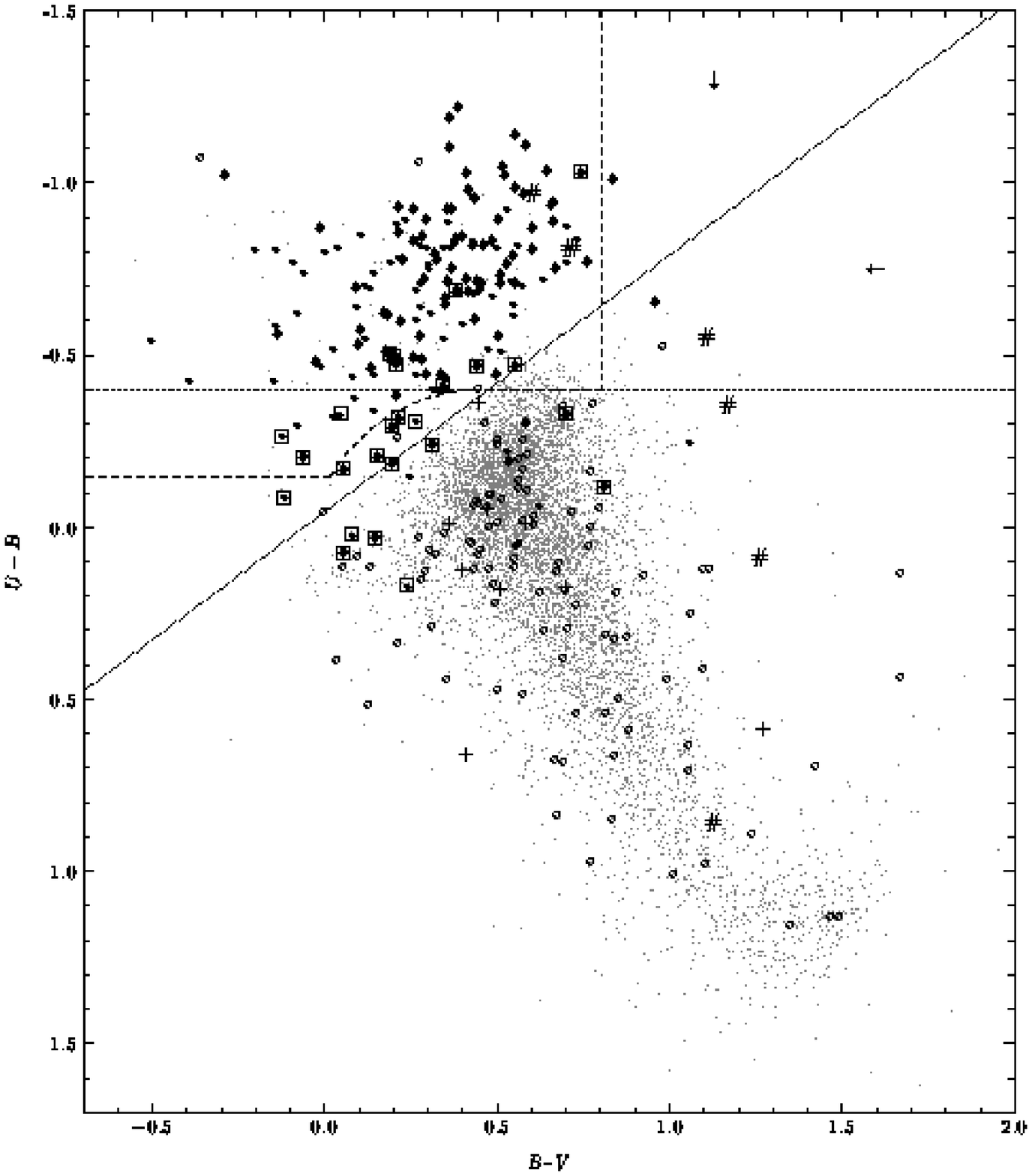}}
\caption{\label{colour_colour}
Colour-colour diagram for the M\,3 field. QSOs/Sey\,1s are shown as
{\tiny$\bullet$} (present study) or {\tiny$\blacklozenge$} (CFHT survey),
respectively. Quasars with $z>2.2$ are framed with a box.
The arrows on the right and at the top, respectively, indicate
QSOs with unknown $B-V$ and $U-B$ colours, respectively. 
The \# symbols indicate narrow emission line galaxies. Open circles are
VPM-QSO candidates that were spectroscopically identified as 
foreground stars; medium-priority candidates without spectroscopic 
follow-up observations are shown as plus signs.
Other star-like objects with $14<B<20$ are shown as
small dots. 
Selection criteria from colour surveys are indicated by 
{\it horizontal dotted line:} UVX search, 
{\it diagonal dotted line:} 
two-colour search as discussed in Paper~1, 
{\it dashed curve:} two-colour selection according to LaFranca et al.
(\cite{LaF92}).
}
\end{figure*}
\afterpage

\begin{figure*}[ht]
\resizebox{17.5cm}{13.0cm}{\includegraphics{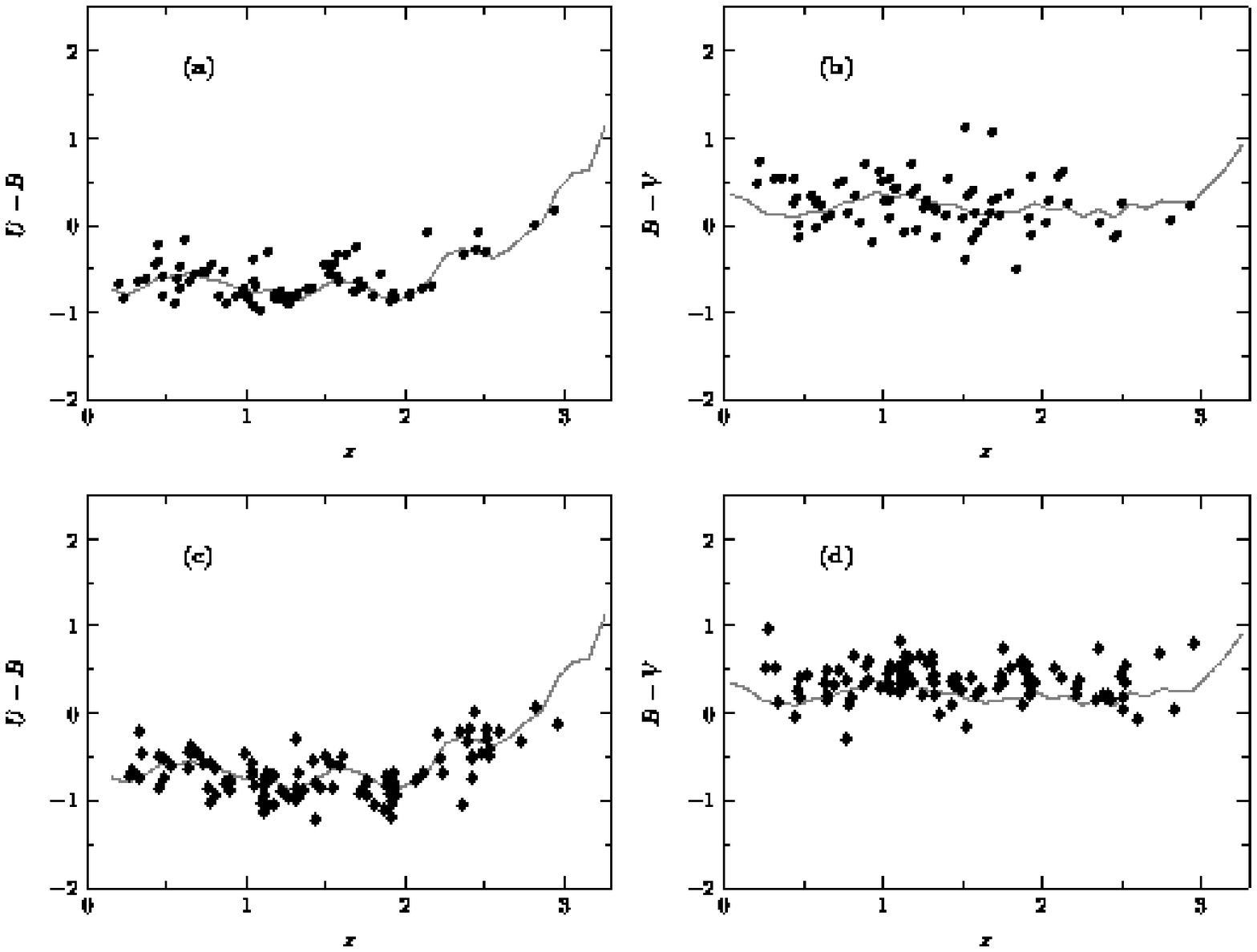}}
\caption{\label{color_z}
Colour-redshift relations for the the QSOs/Sey1s in the M\,3 field. 
Panels {\bf a,b} show the objects from Table\,3
($\bullet$). Panels {\bf c,d} show the QSOs/Sey1s from the NED
identified with objects from our list
({\tiny$\blacklozenge$}).
For comparison, the mean relations for the 
QSOs from V\'eron-Cetty \& V\'eron(\cite{Ver01}) 
are plotted. 
}
\end{figure*}
\afterpage

%
\subsection{\label{spectra}Spectra}
%

The low-resolution spectra of the objects from Table\,3 
are shown in Fig.\,\ref{allspectra}. The spectra are dominated by the typical
AGN-emission lines:
Ly$\alpha$+\ion{N}{v}$\lambda1240$,
\ion{Si}{iv}+\ion{O}{iv}]$\lambda1400$,
\ion{C}{iv}$\lambda1549$,
\ion{C}{iii}]$\lambda1909$,
\ion{Mg}{ii}$\lambda2798$,
[\ion{O}{iii}]$\lambda\lambda4959,5007$,
and the Balmer series.
A few objects (No. 10, 19, 64, 50, 11, 69, 7)  apparently have relatively 
weak lines. Unfortunately, many of these spectra were taken at
relatively bad atmospheric transparency, and poorly 
removed telluric lines may effect the equivalent widths of
the QSO emission lines.
Therefore we do not quantitatively discuss the
distribution of the equivalent widths in this paper. 
The analysis of the QSOs from the VPM survey in the M\,92 field
(where most of the spectra were taken under 
better weather conditions) has shown that the sample-averaged
line equivalent widths for the VPM QSOs are in good agreement
with those from other samples of radio-quiet QSOs
(Meusinger \& Brunzendorf \cite{Meu01}).

Broad absorption troughs are indicated in the spectra of the
QSOs No. 3, 49, 37, 57, 7, 44, 75, 33, 65, and 70. For some other QSOs
absorption features may be hidden due to the low
signal-to-noise. The fraction of BAL QSOs is about 10\%, in good
agreement with the
BAL percentage in the SDSS Early Data Release (Schneider et al.
\cite{Sch02}). There is only one object with an unusual spectrum:
the BAL QSO No.\,7 where the emission lines are almost completely
masked by extremely broad absortion lines.  The best guess for the
emission redshift is $z_{\rm em} \approx 1.7$, compared to
$z_{\rm abs} \approx 1.5$ for the strongest absorption lines
(\ion{C}{iv}$\lambda1549$, \ion{Al}{iii}$\lambda1860$, and the
\ion{Fe}{ii}-multiplet at $\lambda2500$). This object is a high-priority
VPM QSO candidate with quite red colours (see below). There is no entry in
the NED at this position. Objects like No.\,7 are not very
likely to be recognized by most other optical QSO surveys.
For a few other QSOs/Sey1s, the spectra in Fig.\,\ref{allspectra}
have unusually red continua (No. 42, 64, 48, 3).
However, the $U-B$ indices (Table\,3) of these objects
closely follow the mean colour-redshift relation (Fig.\,\ref{color_z})
and thus the missing blue light in the spectra is interpreted by
the slit-loss effect due to atmospheric dispersion (Sect.\,1).
We conclude that, up to the limit of the survey, the fraction of QSOs 
with unusual spectra is at maximum a few percent. This conclusion is 
again in agreement with the statistics from the (still incomplete) SDSS 
data (Hall et al. \cite{Hal02}).

%
\subsection{\label{colours}Colour indices}
%

In Paper\,1, a colour-colour diagram of the QSO candidates was
presented showing a broad scatter of their colour indices and
a large fraction of red QSO candidates. The distribution
of the spectroscopically classified  objects
on the $U-B$ versus $B-V$ plane is shown in Fig.\,\ref{colour_colour}.
The most important result is the fact that all candidates with 
{\it extremely} red colours proved
to be foreground stars, in agreement with what we found
from the VPM survey in the M\,92 field.
Obviously, the QSOs from Table\,3  
populate essentially the same
area as the QSOs from the CFHT grens survey. For $z<2.2$ this area
is well defined by the selection criteria of classical colour
surveys.  There are only 7 low-redshift ($z<2.2$) QSOs located
beyond the demarcation line for colour selection discussed in
Paper\,1. A typical fluctuation of about 0.35\,mag per colour index
is expected due to photometric errors and variability (as the time-lines 
for the three colour bands are not identical). In addition, a 
scatter may be produced by intrinsic differences in the continuum slope 
and/or the strength of emision lines and/or absorption troughs.
Remarkably, the strongest deviation from the colour selection line 
is measured for the two absorption line QSOs No.\,7 and 75. 

The same conclusion is reached from the colour-redshift relations
(Fig.\,\ref{color_z}). Apart from the scatter due to variability and 
photometric errors, the QSOs from our sample closely
follow the mean relation of QSOs from the V\'eron-Cetty \& V\'eron
(2001) catalogue. Among the QSOs from Table\,3, 
the strongest deviation is measured again for Nos.\,7 and 75. The QSO
No.\,28,  which is the faintest object in Table\,3, 
shows a strong deviation in $B-V$. Fainter QSOs tend to have larger 
colour indices $B-V$ (Fig.\,\ref{color_z}d).

Data from the 2 Micron All Sky Survey (2MASS; Skrutskie 
et al. \cite{Skr97}), March 2000 data release are available
for $~25$\% of the field. With an identification radius of 10\,arcsec 
we identified six QSOs from our whole sample with catalogued 2MASS sources. 
For all six
objects the $B-K_{\rm s}$ colour index is smaller than 4, i.\,e.
smaller than for the red QSOs found by Webster et al. (\cite{Web95})
among the flat-spectrum radio-loud QSOs.

For the M\,92 field we have estimated that the fraction of
QSOs with unusually red $B-V$ colour indices must be less than 3\% up
to $B=19.8$ (Brunzendorf \&
Meusinger \cite{Bru02}). From the data in the M\,3 field we
estimate a similar fraction of about 2\% up to $B=19.7$.
Although a survey in the $B$-band is obviously not an ideal
approach to derive strong conclusions about the underlying population of
possibly highly reddened QSOs, the unbiased VPM QSO sample
provides a constraint of its properties. 
Let us assume for simplicity that there are
two QSO populations of comparable size: normal QSOs and reddened QSOs with an
intrinsic dust reddening equivalent to $E_{\rm B-V} = 0.5$ -- a not
unreasonable level in a dusty system -- implying an extinction of about
2\,mag in the $B$-band (assuming a galactic extinction curve). Using
the number-magnitude relation from Fig.\,\ref{fl_dichte}, we would expect to
detect about 5-7 strongly reddened QSOs up to $B_{\rm lim} = 19.7$ in each
search field. This is clearly more than what we found,
indicating that the red QSOs are either redder on average or less frequent. 
A more detailed
discussion of this question has to be deferred to a separate study.

%
\subsection{\label{pairs}QSO pairs}
%

The search for pairs with an angular separation of up to 2\,arcmin
yields 8 combinations, but the redshift differences are
very large for 7 of them. The closest pair in the three-dimensional space 
consists of Nos. 38 and 40 from  Table\,4 
with $z=1.310$ and 1.325 and an angular distance of 96\,arcsec.
For objects with small separations (e.\,g., $<10$\,arcsec),  
the measurements of variability and proper motion are attended
with additional uncertainties. Such objects are, in principle, 
rejected from the VPM candidate list.  This does not significantly
reduce the general efficiency of the search method but the 
efficiency of the detection of close pairs.

%
\section{\label{conclusions}Conclusions}
%

We performed a VPM QSO survey with a limiting magnitude of $B_{\rm lim} \approx 20$  
in a 10 square degrees field at high galactic latitude. The VPM 
technique proved to be an efficient method for finding QSOs.
As the result of the spectroscopic follow-up observations of
198 candidates and the identification of further candidates in the NED,
a sample of 175 QSOs/Sey1s with $0.4<z<3$ is available.
With a stellar contamination of only about 4\%, the high-priority 
QSO subsample from the VPM search is very clean. 
For the combined sample of high-and-medium-priority
objects the fraction of established QSOs/Sey1s among
all candidates is still as high as 63\%.
The completeness of the VPM QSO sample with $B \le 19.7$ 
 is estimated to be
94\%. The number-magnitude relation for that sample is in good agreement 
with the one expected from the relation derived by Wisotzki (\cite{Wis98}) 
from various QSO samples. At brighter magnitudes ($B<18.5$), we find a
somewhat higher QSO surface density.

The optical broad-band colours and the spectra of the
VPM-selected QSOs are not significantly different from those of QSOs
selected by other optical surveys, in agreement with what we found
in the M\,92 field (Meusinger \& Brunzendorf \cite{Meu01};
Brunzendorf \& Meusinger \cite{Bru02}).
Such a result can not be a priori expected since the selection criteria of the
VPM survey are completely different from those in 
most other optical surveys.
Although there is a large fraction of objects with red colours among the
VPM QSO candidates, all candidates with extremely red colours were proved to be
stellar contaminants. We estimate that the fraction of QSOs with unusualy
red optical colours is at most a few per cent up to the limit of the survey,
provided that their variability properties are not significantly different from those
of the other QSOs. Some BAL QSOs are known to be considerably redder
than the targets of most QSO surveys (e.\,g.,
Weymann et al. \cite{Wey91};
Menou et al. \cite{Men01};
Hall et al. \cite{Hal02}).
The fraction of such unusual QSOs
in the (incomplete) SDSS Early Data Release is less than 1\%
(Hall et al. \cite{Hal02}), in good agreement with our result.
In this context it is notable that all VPM QSOs with indication
for substantial absorption
are strongly variabel ($I_{\rm var} > 1.5$). A VPM search is thus
expected to be essentially unbiased against strongly absorbed QSOs,
apart from the bias introduced by the band-pass of the search.
The general agreement of the properties of the  VPM QSO sample with
those from more conventional optical surveys
suggests that the latter  do obviously
not ignore a substantial number of red QSOs up to $B \approx 20$. On the
other hand, we can conclude that the VPM survey can be combined with
colour search criteria in order to reach a very high efficiency without
a significant loss of  completeness. Of course, we can not exclude the
existence of substantial numbers of obscured red QSOs that are fainter
than the current survey limit. Such objects can be found by a deeper
VPM survey.



\begin{table*}[ht]
{\renewcommand{\baselinestretch}{0.95}\footnotesize
\begin{tabular}{rlllrrrrrrrrr}
\toprule
No. & $\alpha$\,(J2000) & $\delta$\,(J2000) & \multicolumn{1}{c}{$z$} & run & $M_B$ & type & \multicolumn{1}{c}{$B$} 
& $U-B$ & $B-V$ & $I_{\rm pm}$ & $I_{\rm var}$ &
$I_{\rm ltvar}$\\
\midrule
%
 1 &  13 35 14.55 & 29 44 55.3 &  1.853 &  4  & -26.39 & QSO & 19.49 & -0.54 & -0.50  & 2.33 & 1.62  &  3.57\\
 2 &  13 35 42.48 & 28 53 30.5 &  0.647 &  2  & -24.73 & QSO & 18.40 & -0.64 &  0.09  & 1.38 & 3.19  &  6.36\\
 3 &  13 35 50.27 & 28 48 08.7 &  1.071 &  1  & -26.94 & QSO & 17.49 & -0.68 &  0.43  & 3.32 & 3.34  &  8.13\\
 4 &  13 36 09.57 & 29 09 16.3 &  0.560 &  5  & -23.23 & QSO & 19.55 & -0.89 &  0.36  & 1.93 & 1.69  &  1.82\\
 5 &  13 36 10.89 & 27 10 52.8 &  1.718 &  2  & -26.95 & QSO & 18.69 & -0.62 &  0.31  & 1.67 & 3.05  &  6.14\\
 6 &  13 36 15.06 & 28 59 11.9 &  1.419 &  4  & -25.78 & QSO & 19.37 & -0.71 &  0.54  & 1.27 & 1.95  &  3.40\\
 7 &  13 37 51.14 & 27 14 21.1 &  1.70: &  5  & -26.09 & QSO & 19.52 & -0.26 &  1.06  & 0.22 & 2.51  &  2.47\\
 8 &  13 37 56.25 & 29 38 39.6 &  1.216 &  3  & -25.45 & QSO & 19.28 & -0.83 &  0.43  & 1.32 & 2.77  &  2.12\\
 9 &  13 37 59.79 & 28 17 05.0 &  0.888 &  3  & -24.83 & QSO & 19.14 & -0.87 &  0.71  & 1.69 & 2.37  &  4.45\\
10 &  13 38 14.68 & 28 46 12.7 &  0.376 &  4  & -22.65 &Sey1 & 19.23 & -0.62 &  0.55  & 0.50 & 2.49  &  8.56\\
11 &  13 38 26.47 & 28 36 37.8 &  1.534 &  3  & -26.06 & QSO & 19.28 & -0.55 &  0.35  & 0.54 & 1.65  &  2.09\\
12 &  13 38 44.36 & 29 01 49.2 &  1.052 &  2  & -26.02 & QSO & 18.37 & -0.64 &  0.28  & 1.94 & 3.48  & 10.00\\
13 &  13 38 48.93 & 28 54 59.0 &  1.281 &  4  & -25.38 & QSO & 19.50 & -0.89 &  0.23  & 1.00 & 1.94  &  3.39\\
14 &  13 38 52.81 & 29 33 35.3 &  1.322 &  4  & -25.61 & QSO & 19.35 & -0.78 &  0.22  & 2.33 & 1.91  &  3.55\\
15 &  13 39 01.52 & 29 19 11.0 &  0.137:&  5  & -21.61 &NELG & 18.06 &  0.86 &  1.13  & 7.11 & 1.22  &	  *\\
16 &  13 39 13.30 & 27 18 18.5 &  0.680 &  3  & -24.25 & QSO & 19.01 & -0.55 &  0.12  & 0.79 & 1.65  &  1.81\\
17 &  13 39 18.46 & 29 29 52.4 &  1.743 &  4  & -26.20 & QSO & 19.49 & -0.70 &  0.13  & 2.45 & 2.40  &  3.10\\
18 &  13 39 19.27 & 28 39 08.9 &  2.514 &  3  & -27.86 & QSO & 18.98 & -0.31 &  0.27  & 0.69 & 1.91  &  3.74\\
19 &  13 39 58.17 & 29 37 54.0 &  0.441 &  6  & -22.73 &Sey1 & 19.48 & -0.44 &  0.27  & 2.68 & 1.69  &  2.43\\
20 &  13 40 11.33 & 29 24 00.5 &  1.527 &  4  & -25.92 & QSO & 19.41 & -0.43 & -0.39  & 1.80 & 6.06  & 34.40\\
21 &  13 40 33.34 & 29 23 16.4 &  0.841 &  2  & -25.11 & QSO & 18.72 & -0.81 &  0.34  & 1.01 & 3.00  &  4.19\\
22 &  13 41 04.99 & 29 17 08.0 &  1.100 &  3  & -25.17 & QSO & 19.32 & -0.96 &  0.43  & 2.25 & 2.09  &  2.93\\
23 &  13 41 05.08 & 28 38 29.7 &  0.580 &  5  & -23.50 & QSO & 19.36 & -0.61 &  0.26  & 1.68 & 1.71  &  1.87\\
24 &  13 41 07.37 & 28 39 35.7 &  1.569 &  4  & -26.50 & QSO & 18.89 & -0.59 &  0.40  & 0.30 & 2.80  &  5.79\\
25 &  13 41 29.81 & 29 42 31.0 &  0.330 &  3  & -22.61 &Sey1 & 19.00 & -0.65 &  0.55  & 1.65 & 1.87  &  1.96\\
26 &  13 42 00.57 & 29 47 31.9 &  0.190 &  5  & -20.86 &NELG & 19.53 &  0.09 &  1.27  & 1.07 & 2.45  &  1.28\\
27 &  13 42 01.08 & 29 00 33.4 &  1.016 &  4  & -24.93 & QSO & 19.38 & -0.82 &  0.28  & 0.46 & 2.56  &  7.18\\
28 &  13 42 08.01 & 29 44 34.7 &  1.528 &  5  & -24.90 & QSO & 20.42 &     * &  1.13  & 3.32 & 0.95  &     *\\
29 &  13 42 08.28 & 27 09 31.0 &  1.190 &  3  & -26.55 & QSO & 18.13 & -0.82 &  0.38  & 0.67 & 3.21  & 14.90\\
30 &  13 42 29.81 & 28 42 48.9 &  1.184 &  5  & -24.93 & QSO & 19.74 & -0.77 &  0.70  & 1.31 & 3.49  &  6.29\\
31 &  13 42 30.41 & 28 56 26.4 &  1.264 &  4  & -25.34 & QSO & 19.50 & -0.88 &  0.21  & 0.83 & 1.88  &  4.06\\
32 &  13 42 30.90 & 28 37 25.1 &  0.589 &  3  & -24.75 & QSO & 18.14 & -0.71 &  0.28  & 1.71 & 3.42  &  9.15\\
33 &  13 42 32.55 & 28 55 19.3 &  2.172 &  5  & -26.77 & QSO & 19.58 & -0.69 &  0.27  & 2.13 & 2.50  &  3.14\\
34 &  13 42 39.19 & 29 36 13.4 &  0.235 &  3  & -21.48 &Sey1 & 19.36 & -0.84 &  0.73  & 2.80 & 5.66  & 10.70\\
35 &  13 42 53.05 & 29 33 42.8 &  1.939 &  4  & -26.66 & QSO & 19.37 & -0.82 &  0.56  & 2.00 & 2.39  &  1.85\\
36 &  13 42 58.55 & 27 54 31.2 &  1.799 &  2  & -26.91 & QSO & 18.88 & -0.81 &  0.36  & 1.31 & 2.78  &  4.44\\
37 &  13 43 02.76 & 28 49 43.8 &  1.581 &  2  & -26.57 & QSO & 18.83 & -0.34 &  0.14  & 2.05 & 1.73  &  1.80\\
38 &  13 43 05.17 & 29 09 21.2 &  1.679 &  5  & -26.00 & QSO & 19.58 & -0.74 &  0.15  & 2.14 & 3.17  &  9.97\\
39 &  13 43 24.63 & 29 39 07.1 &  1.332 &  3  & -25.83 & QSO & 19.15 & -0.81 & -0.14  & 0.78 & 2.17  &  4.09\\
40 &  13 43 47.46 & 28 35 07.3 &  0.477 &  4  & -22.97 & QSO & 19.43 & -0.80 &  0.00  & 1.86 & 3.25  &  6.87\\
41 &  13 43 48.08 & 28 23 53.2 &  0.733 &  3  & -24.28 & QSO & 19.18 & -0.52 &  0.47  & 2.29 & 2.30  &  3.73\\
42 &  13 43 50.03 & 28 02 05.7 &  0.211 &  1  & -22.69 &Sey1 & 17.92 & -0.67 &  0.48  & 1.53 & 3.81  &  5.24\\
43 &  13 43 58.99 & 28 35 48.8 &  0.193 &  4  & -21.04 &NELG & 19.38 & -0.55 &  1.11  & 1.33 & 1.83  &  2.34\\
44 &  13 43 59.43 & 29 29 33.3 &  2.109 &  5  & -26.85 & QSO & 19.44 & -0.71 &  0.58  & 1.47 & 1.51  &  1.08\\
45 &  13 44 03.07 & 29 20 05.3 &  1.506 &  4  & -25.79 & QSO & 19.50 & -0.43 &  0.08  & 0.53 & 4.19  &  4.65\\
46 &  13 44 12.71 & 28 52 16.0 &  2.375 &  2  & -28.20 & QSO & 18.40 & -0.33 &  0.05  & 0.94 & 1.36  &  2.34\\
47 &  13 44 22.39 & 28 56 42.9 &  0.481 &  3  & -23.07 & QSO & 19.35 & -0.58 & -0.14  & 1.50 & 4.21  &  9.53\\
48 &  13 44 33.95 & 28 27 23.7 &  1.007 &  3  & -24.93 & QSO & 19.36 & -0.81 &  0.50  & 0.92 & 2.48  &  2.69\\
49 &  13 45 04.05 & 28 48 18.6 &  1.275 &  3  & -25.59 & QSO & 19.27 & -0.81 &  0.30  & 2.59 & 1.73  &  1.98\\
50 &  13 45 11.65 & 28 13 60.0 &  1.053 &  5  & -24.71 & QSO & 19.68 & -0.92 &  0.53  & 1.85 & 1.89  &  3.00\\
51 &  13 45 13.81 & 29 06 03.0 &  0.587 &  2  & -24.53 & QSO & 18.36 & -0.47 & -0.01  & 0.82 & 4.19  & 11.00\\
52 &  13 45 38.38 & 28 49 35.5 &  0.454 &  2  & -23.56 & QSO & 18.74 & -0.22 &  0.53  & 0.59 & 1.17  &  1.27\\
53 &  13 45 47.42 & 28 19 09.8 &  0.457 &  4  & -22.91 & QSO & 19.39 & -0.40 &  0.32  & 2.71 & 1.72  &  1.81\\
54 &  13 45 57.82 & 28 32 06.7 &  1.700 &  2  & -26.64 & QSO & 18.97 & -0.75 &  0.30  & 1.08 & 2.06  &  3.09\\
55 &  13 46 22.15 & 28 52 14.4 &  0.758 &  5  & -23.77 & QSO & 19.78 & -0.51 &  0.51  & 1.12 & 3.21  &  3.82\\
\bottomrule			       
\end{tabular}		 	       
\caption{\label{QSO_list}											       
QSOs, Seyfert\,1, and NELGs from the follow-up spectroscopy 							       
of the present study. $I_{\rm pm},\, I_{\rm var},\, I_{\rm ltvar}$ are
the indices for proper motion, overall variability, and long-term variability,
respectively. 
A colon behind the redshift symbolises uncertain data, an asterisk indicates
missing data. (Note that $I_{\rm ltvar}$ has been computed only for objects with 
$B<20$ and $I_{\rm pm}<4$.)}												       
}														       
\end{table*}													       
														       
\begin{table*}[ht]													       
{\renewcommand{\baselinestretch}{0.95}\footnotesize								       
\begin{tabular}{rrrrrrrrrrrrr}											       
\toprule													       
No. & $\alpha$\,(J2000) & $\delta$\,(J2000) & \multicolumn{1}{c}{$z$} & run & $M_B$ & type & 				       
\multicolumn{1}{c}{$B$} & $U-B$ & $B-V$ & $I_{\rm pm}$ & $I_{\rm var}$ &					       
$I_{\rm ltvar}$\\											       
\midrule
56 &  13 46 38.31 & 29 45 54.7 &  1.336 &  4  & -25.58 & QSO & 19.41 & -0.77 &  0.19  & 1.06 & 1.93  &  4.83\\
57 &  13 46 48.04 & 29 46 21.0 &  1.594 &  5  & -25.90 & QSO & 19.52 & -0.62 & -0.08  & 0.68 & 1.88  &  2.43\\
58 &  13 47 04.33 & 29 35 23.3 &  0.625 &  2  & -24.40 & QSO & 18.65 & -0.15 &  0.25  & 1.41 & 1.55  &  2.87\\
59 &  13 47 14.11 & 28 00 08.0 &  1.566 &  3  & -26.09 & QSO & 19.29 & -0.43 & -0.14  & 2.97 & 1.59  &  2.47\\
60 &  13 47 15.42 & 29 26 23.9 &  1.918 &  2  & -27.25 & QSO & 18.75 & -0.85 &  0.11  & 1.14 & 3.50  &  4.54\\
61 &  13 47 23.48 & 29 27 22.2 &  1.223 &  2  & -25.78 & QSO & 18.97 & -0.74 & -0.06  & 1.93 & 2.07  &  3.83\\
62 &  13 47 35.79 & 29 00 12.3 &  1.398 &  3  & -25.85 & QSO & 19.23 & -0.71 &  0.12  & 0.65 & 1.56  &  2.40\\
63 &  13 47 37.92 & 27 14 50.5 &  0.787 &  5  & -24.07 & QSO & 19.58 & -0.44 &  0.14  & 2.01 & 2.78  &  5.67\\
64 &  13 47 42.88 & 28 56 52.1 &  0.985 &  3  & -24.97 & QSO & 19.28 & -0.72 &  0.64  & 1.85 & 1.95  &  3.76\\
65 &  13 47 56.49 & 28 35 20.7 &  2.458 &  3  & -27.75 & QSO & 18.98 & -0.27 & -0.12  & 0.55 & 2.92  &  3.20\\
66 &  13 47 58.90 & 27 24 50.7 &  1.048 &  2  & -26.16 & QSO & 18.22 & -0.38 &  0.09  & 3.35 & 2.12  &  2.10\\
67 &  13 48 02.84 & 28 27 15.9 &  0.864 &  2  & -25.57 & QSO & 18.32 & -0.52 &  0.03  & 0.50 & 2.51  &  3.22\\
68 &  13 48 04.29 & 28 40 25.0 &  2.471 &  1  & -28.91 & QSO & 17.85 & -0.08 & -0.11  & 1.33 & 3.32  &  6.88\\
69 &  13 48 08.94 & 28 02 13.4 &  1.643 &  1  & -27.55 & QSO & 17.96 & -0.32 &  0.03  & 0.88 & 1.63  &  2.00\\
70 &  13 48 11.70 & 28 18 01.5 &  2.941 &  1  & -29.87 & QSO & 17.72 &  0.17 &  0.24  & 2.22 & 1.81  &  1.14\\
71 &  13 48 14.87 & 29 02 56.3 &  0.938 &  3  & -24.76 & QSO & 19.35 & -0.81 & -0.20  & 2.26 & 4.28  & 11.50\\
72 &  13 48 18.75 & 29 10 41.9 &  1.932 &  4  & -26.79 & QSO & 19.24 & -0.77 & -0.09  & 0.90 & 3.11  &  4.56\\
73 &  13 48 20.29 & 29 34 03.5 &  2.041 &  6  & -26.56 & QSO & 19.65 & -0.76 &  0.30  & 1.41 & 2.26  &  2.07\\
74 &  13 48 22.58 & 28 39 43.2 &  1.140 &  4  & -25.36 & QSO & 19.21 & -0.29 & -0.08  & 1.02 & 2.23  &  3.59\\
75 &  13 48 24.33 & 28 32 50.3 &  2.141 &  5  & -26.76 & QSO & 19.57 & -0.06 &  0.62  & 0.94 & 1.98  &  2.08\\
76 &  13 48 26.47 & 28 43 17.2 &  2.031 &  5  & -27.47 & QSO & 18.72 & -0.81 &  0.04  & 3.53 & 1.67  &  1.26\\
77 &  13 48 26.60 & 29 06 22.7 &  2.819 &  3  & -28.17 & QSO & 19.18 &  0.02 &  0.08  & 1.22 & 1.85  &  1.77\\
\bottomrule
\end{tabular}	
\addtocounter{table}{-1}	 
\caption{
continued}
}
\end{table*}
\afterpage

\begin{table*}[htbp]
{\renewcommand{\baselinestretch}{0.95}\footnotesize
\begin{tabular}{rlllrrrrrrrrr}
\toprule
No. & $\alpha$\,(J2000) & $\delta$\,(J2000) & \multicolumn{1}{c}{$z$} 
& $M_B$ & type & \multicolumn{1}{c}{$B$} 
& $U-B$ & $B-V$ & $I_{\rm pm}$ & $I_{\rm var}$ &
$I_{\rm ltvar}$\\
\midrule
  1 & 13 35 11.82 & 27 56 51.7 & 2.425 &  -26.83  & QSO & 19.85 & -0.74 &      * &   2.07 &   2.02 &  3.33\\
  2 & 13 35 23.65 & 28 08 38.9 & 0.904 &  -24.15  & QSO & 19.87 & -0.87 &   0.60 &   1.60 &   1.80 &  2.38\\
  3 & 13 35 24.10 & 27 17 21.0 & 0.879 &  -24.21  & QSO & 19.74 & -0.78 &   0.33 &   3.61 &   1.14 &  2.65\\
  4 & 13 35 35.92 & 27 21 23.8 & 0.783 &  -24.55  & QSO & 19.08 & -0.58 &   0.10 &   1.97 &   2.54 &  1.88\\
  5 & 13 35 39.68 & 28 05 04.7 & 1.095 &  -24.22  & QSO & 20.26 & -1.02 &   0.52 &   1.55 &   1.36 &	 *\\
  6 & 13 35 46.10 & 28 05 51.1 & 2.960 &  -28.30  & QSO & 19.32 & -0.12 &   0.81 &   1.97 &   1.53 &  1.29\\
  7 & 13 35 48.08 & 27 28 35.5 & 1.331 &  -25.10  & QSO & 19.88 & -0.96 &   0.44 &   2.71 &   1.82 &  1.73\\
  8 & 13 35 54.38 & 27 53 23.2 & 1.886 &  -26.64  & QSO & 19.30 & -0.81 &   0.60 &   1.79 &   1.32 &  2.42\\
  9 & 13 36 02.73 & 27 27 51.8 & 1.117 &  -24.78  & QSO & 19.74 & -1.10 &   0.37 &   2.46 &   2.23 &  4.04\\
 10 & 13 36 05.91 & 28 13 24.1 & 2.388 &  -27.58  & QSO & 19.04 & -0.32 &   0.21 &   2.92 &   1.54 &  1.52\\
 11 & 13 36 11.38 & 27 10 13.0 & 2.425 &  -28.74  & QSO & 17.94 & -0.47 &   0.21 &   3.52 &   1.68 &  2.71\\
 12 & 13 36 13.21 & 28 24 58.6 & 1.908 &  -27.17  & QSO & 18.81 & -1.19 &   0.36 &   1.18 &   1.24 &  1.82\\
 13 & 13 36 16.62 & 28 04 52.8 & 0.271 &  -21.81  & Sey1& 19.35 & -0.71 &   0.51 &   2.15 &   1.47 &  2.51\\
 14 & 13 36 23.02 & 27 07 20.2 & 1.931 &  -26.14  & QSO & 19.88 & -0.78 &   0.22 &   1.02 &   1.85 &  3.78\\
 15 & 13 36 30.17 & 27 01 01.3 & 0.283 &  -21.08  & NELG& 20.17 & -0.81 &   0.71 &   1.79 &   1.66 &	 *\\
 16 & 13 36 43.50 & 27 29 53.6 & 0.780 &  -23.83  & QSO & 19.56 & -1.02 &  -0.29 &   2.21 &   3.88 & 41.80\\
 17 & 13 36 43.59 & 27 10 59.3 & 1.440 &  -25.02  & QSO & 20.17 & -1.22 &   0.39 &   1.74 &   2.67 &	 *\\
 18 & 13 36 50.06 & 27 20 54.8 & 1.950 &  -26.11  & QSO & 19.94 & -0.93 &   0.37 &   1.06 &   1.01 &  1.21\\
 19 & 13 36 56.79 & 27 25 43.0 & 1.360 &  -26.21  & QSO & 18.83 & -0.87 &  -0.01 &   3.86 &   4.89 &  9.90\\
 20 & 13 37 00.88 & 27 13 22.8 & 2.074 &  -26.27  & QSO & 19.98 & -0.76 &   0.53 &   0.94 &   1.95 &  3.25\\
 21 & 13 37 14.12 & 27 12 49.8 & 1.909 &  -26.47  & QSO & 19.51 & -0.99 &   0.55 &   1.02 &   1.68 &  1.50\\
 22 & 13 37 17.39 & 27 01 07.4 & 0.637 &  -24.45  & QSO & 18.64 & -0.43 &   0.35 &   2.41 &   3.02 &  3.02\\
 23 & 13 37 19.98 & 28 09 26.2 & 0.460 &  -22.69  & Sey1& 19.62 & -0.84 &   0.28 &   1.38 &   1.70 &  2.80\\
 24 & 13 37 26.09 & 27 36 39.1 & 1.121 &  -25.21  & QSO & 19.32 & -0.72 &   0.29 &   1.22 &   1.59 &  3.84\\
 25 & 13 37 35.34 & 27 03 11.2 & 1.762 &  -25.58  & QSO & 20.14 & -0.77 &   0.76 &   1.31 &   2.46 &	 *\\
 26 & 13 37 37.95 & 28 13 47.1 & 1.865 &  -25.72  & QSO & 20.13 & -1.11 &   0.58 &   1.66 &   1.51 &	 *\\
 27 & 13 37 39.07 & 28 04 46.8 & 1.124 &  -24.24  & QSO & 20.30 & -1.14 &   0.55 &   1.96 &   1.60 &	 *\\
 28 & 13 37 44.30 & 27 01 29.2 & 1.928 &  -27.21  & QSO & 18.81 & -0.84 &   0.26 &   3.79 &   3.56 &  8.39\\
 29 & 13 37 49.44 & 28 04 01.0 & 1.321 &  -25.19  & QSO & 19.77 & -0.86 &   0.21 &   2.48 &   2.08 &  2.01\\
 30 & 13 37 54.57 & 28 09 44.0 & 1.592 &  -26.32  & QSO & 19.10 & -0.60 &   0.22 &   1.39 &   1.96 &  2.31\\
 31 & 13 37 54.90 & 27 37 55.6 & 2.362 &  -26.34  & QSO & 20.24 & -1.03 &   0.74 &   1.47 &   1.90 &	 *\\
 32 & 13 37 55.12 & 27 22 44.4 & 0.433 &  -21.69  & NELG& 20.48 & -0.97 &   0.60 &   1.23 &   1.29 &	 *\\
 33 & 13 38 04.56 & 27 48 34.4 & 0.814 &  -23.33  & QSO & 20.41 & -0.95 &   0.66 &   2.47 &   1.40 &	 *\\
 34 & 13 38 21.80 & 29 14 45.7 & 0.646 &  -26.02  & QSO & 17.11 & -0.62 &   0.17 &   1.50 &   2.19 &  6.49\\
 35 & 13 38 29.51 & 28 02 56.4 & 1.116 &  -24.14  & QSO & 20.38 & -1.01 &   0.84 &   1.80 &   1.25 &	 *\\
 36 & 13 38 34.08 & 27 30 09.8 & 0.342 &  -21.89  & Sey1& 19.80 & -0.47 &   0.14 &   0.87 &   1.72 &  4.46\\
 37 & 13 38 49.26 & 27 49 15.7 & 1.310 &  -24.71  & QSO & 20.23 & -0.94 &   0.66 &   0.51 &   0.96 &	 *\\
 38 & 13 38 55.70 & 27 13 05.3 & 0.486 &  -22.42  & Sey1& 20.02 & -0.52 &   0.18 &   1.42 &   1.94 &	 *\\
 39 & 13 38 55.91 & 27 48 38.4 & 1.325 &  -25.72  & QSO & 19.25 & -0.67 &   0.35 &   1.21 &   2.88 &  5.78\\
 40 & 13 38 57.86 & 27 11 50.3 & 1.922 &  -25.70  & QSO & 20.30 & -1.03 &   0.41 &   0.38 &   0.68 &	 *\\
 41 & 13 39 00.98 & 28 14 24.8 & 2.513 &  -27.43  & QSO & 19.40 & -0.17 &   0.06 &   1.89 &   1.68 &  3.14\\
 42 & 13 39 02.19 & 27 36 43.2 & 2.530 &  -27.11  & QSO & 19.76 & -0.47 &   0.56 &   0.89 &   1.84 &  2.86\\
 43 & 13 39 09.63 & 28 05 26.5 & 1.113 &  -25.29  & QSO & 19.23 & -0.93 &   0.26 &   0.42 &   1.97 &  2.22\\
 44 & 13 39 16.57 & 27 28 16.1 & 1.047 &  -25.16  & QSO & 19.22 & -0.71 &   0.55 &   1.95 &   2.61 &  4.25\\
 45 & 13 39 24.84 & 27 23 35.8 & 1.056 &  -24.50  & QSO & 19.90 & -0.83 &   0.49 &   1.25 &   1.01 &  1.20\\
 46 & 13 39 38.80 & 27 15 29.3 & 1.750 &  -25.22  & QSO & 20.48 & -0.82 &   0.43 &   0.63 &   1.19 &	 *\\
 47 & 13 39 45.89 & 27 11 00.4 & 1.120 &  -24.65  & QSO & 19.88 & -0.82 &   0.46 &   2.61 &   1.71 &  4.83\\
 48 & 13 39 55.56 & 27 47 58.7 & 0.657 &  -24.06  & QSO & 19.11 & -0.45 &   0.49 &   1.17 &   1.35 &  2.00\\
 49 & 13 40 00.10 & 28 14 04.4 & 1.760 &  -25.74  & QSO & 19.98 & -0.93 &   0.36 &   1.43 &   1.42 &  1.60\\
 50 & 13 40 04.87 & 28 16 53.2 & 2.517 &  -29.45  & QSO & 17.39 & -0.29 &   0.20 &   3.58 &   1.29 &  2.88\\
 51 & 13 40 09.35 & 27 18 39.6 & 0.327 &  -21.92  & Sey1& 19.66 & -0.74 &   0.51 &   1.52 &   2.03 &  1.77\\
 52 & 13 40 13.61 & 28 15 20.7 & 1.467 &  -25.35  & QSO & 19.88 & -0.85 &   0.40 &   3.10 &   1.92 &  2.26\\
 53 & 13 40 20.41 & 27 20 41.7 & 1.140 &  -24.63  & QSO & 19.94 & -0.76 &   0.67 &   1.26 &   1.70 &  0.59\\
 54 & 13 40 22.78 & 27 40 58.7 & 0.172 &  -21.19  & NELG& 18.98 & -0.81 &   0.72 &   1.67 &   7.43 & 12.60\\
 55 & 13 40 31.59 & 27 05 41.9 & 0.280 &  -21.59  & Sey1& 19.65 & -0.66 &   0.96 &   1.40 &   0.96 &  0.97\\
\bottomrule			       
\end{tabular}		 	       
\caption{\label{QSO_NED_list}
As Table\,\ref{QSO_list} but for the objects from our basic candidate sample 
that were identified with											       
QSOs, Seyfert\,1s, or NELGs from the NED.}												       
}														       
\end{table*}													       
														       
\begin{table*}[ht]													       
{\renewcommand{\baselinestretch}{0.95}\footnotesize								       
\begin{tabular}{rrrrrrrrrrrrr}											       
\toprule													       
No. & $\alpha$\,(J2000) & $\delta$\,(J2000) & \multicolumn{1}{c}{$z$} & 
$M_B$ &type & 				       
\multicolumn{1}{c}{$B$} & $U-B$ & $B-V$ & $I_{\rm pm}$ & $I_{\rm var}$ &					       
$I_{\rm ltvar}$\\											       
\midrule 													     
%
 56 & 13 40 48.23 & 27 40 09.3 & 1.900  &  -25.57  & QSO & 20.39 & -0.90 &   0.51 &   1.48 &   1.47 &	  * \\
 57 & 13 40 54.50 & 27 17 53.7 & 1.252  &  -24.80  & QSO & 20.02 & -0.93 &   0.21 &   1.43 &   2.13 &	  * \\
 58 & 13 41 00.11 & 27 25 17.9 & 1.175  &  -24.19  & QSO & 20.46 & -1.04 &   0.65 &   1.27 &   1.87 &	  * \\
 59 & 13 41 16.27 & 28 16 09.2 & 1.310  &  -25.03  & QSO & 19.91 & -0.30 &   0.58 &   0.53 &   1.20 &  2.28 \\
 60 & 13 41 19.35 & 27 29 26.7 & 0.480  &  -22.29  & QSO & 20.12 & -0.72 &   0.41 &   0.80 &   1.87 &	  * \\
 61 & 13 41 23.26 & 27 49 55.3 & 1.045  &  -25.36  & QSO & 19.01 & -0.56 &   0.28 &   4.24 &   2.94 &	  * \\
 62 & 13 41 53.82 & 27 35 56.2 & 1.552  &  -25.03  & QSO & 20.33 & -0.85 &   0.40 &   0.24 &   1.24 &	  * \\
 63 & 13 42 03.24 & 26 58 46.7 & 0.530  &  -23.18  & QSO & 19.47 & -0.60 &   0.44 &   2.30 &   1.46 &  1.75 \\
 64 & 13 42 11.61 & 28 28 49.8 & 0.330  &  -22.29  & Sey1& 19.31 & -0.19 &   0.53 &   1.13 &   0.84 &  1.31 \\
 65 & 13 42 14.31 & 27 37 55.0 & 0.770  &  -23.68  & QSO & 19.90 & -0.84 &   0.38 &   1.44 &   1.17 &  0.89 \\
 66 & 13 42 21.55 & 27 26 21.4 & 2.240  &  -26.95  & QSO & 19.47 & -0.69 &   0.38 &   0.40 &   1.52 &  1.50 \\
 67 & 13 42 23.62 & 27 04 34.6 & 2.827  &  -27.63  & QSO & 19.75 &  0.07 &   0.06 &   1.57 &   1.37 &  2.61 \\
 68 & 13 42 24.34 & 27 20 40.0 & 0.704  &  -25.29  & QSO & 18.05 & -0.45 &   0.34 &   0.85 &   1.90 &  2.32 \\
 69 & 13 42 37.30 & 27 15 41.4 & 1.229  &  -24.34  & QSO & 20.43 & -0.89 &   0.66 &   1.70 &   2.24 &	  * \\
 70 & 13 42 44.42 & 27 37 37.2 & 1.721  &  -25.65  & QSO & 20.00 & -0.89 &   0.29 &   2.03 &   1.65 &	  * \\
 71 & 13 42 52.90 & 27 36 15.8 & 1.444  &  -25.13  & QSO & 20.07 & -0.79 &   0.32 &   0.54 &   2.46 &	  * \\
 72 & 13 42 54.02 & 27 33 10.0 & 0.810  &  -24.72  & QSO & 19.01 & -0.62 &   0.19 &   0.68 &   2.02 &  3.16 \\
 73 & 13 42 54.30 & 28 28 05.7 & 1.037  &  -25.98  & QSO & 18.37 & -0.65 &   0.35 &   1.57 &   2.31 &  3.55 \\
 74 & 13 42 54.72 & 26 58 04.9 & 2.737  &  -27.43  & QSO & 19.79 & -0.33 &   0.70 &   3.21 &   1.47 &  0.92 \\
 75 & 13 42 55.06 & 27 53 31.5 & 1.527  &  -26.18  & QSO & 19.14 & -0.56 &  -0.14 &   0.98 &   3.03 &  8.27 \\
 76 & 13 42 59.18 & 27 47 23.1 & 1.164  &  -25.13  & QSO & 19.50 & -0.71 &   0.44 &   1.72 &   4.99 &  3.89 \\
 77 & 13 43 00.10 & 28 44 07.4 & 0.905  &  -26.88  & QSO & 17.14 & -0.75 &   0.37 &   1.87 &   2.55 &  3.94 \\
 78 & 13 43 14.11 & 27 38 12.3 & 2.491  &  -26.67  & QSO & 20.12 & -0.47 &   0.44 &   2.63 &   0.82 &	  * \\
 79 & 13 43 23.05 & 26 57 16.5 & 1.274  &  -25.54  & QSO & 19.32 & -0.97 &   0.58 &   3.28 &   2.18 &  3.04 \\
 80 & 13 43 30.20 & 27 44 55.3 & 2.424  &  -26.56  & QSO & 20.12 & -0.51 &   0.19 &   0.31 &   1.51 &	  * \\
 81 & 13 44 05.32 & 27 26 33.4 & 1.312  &  -25.13  & QSO & 19.81 & -0.98 &   0.42 &   1.39 &   2.24 &  1.79 \\
 82 & 13 44 05.78 & 28 00 04.7 & 0.733  &  -23.70  & QSO & 19.75 & -0.55 &   0.50 &   1.36 &   1.07 &  1.34 \\
 83 & 13 44 26.34 & 27 58 45.0 & 0.900  &  -24.24  & QSO & 19.77 & -0.79 &   0.55 &   1.02 &   1.83 &  1.50 \\
 84 & 13 44 51.91 & 28 11 13.8 & 2.401  &  -27.61  & QSO & 19.03 & -0.18 &   0.20 &   0.54 &   1.47 &  1.66 \\
 85 & 13 45 15.59 & 27 26 17.3 & 1.884  &  -26.71  & QSO & 19.23 & -0.70 &   0.09 &   2.25 &   1.84 &  3.14 \\
 86 & 13 45 40.92 & 27 55 24.6 & 0.453  &  -23.14  & QSO & 19.15 & -0.48 &  -0.03 &   2.41 &   2.58 &  8.77 \\
 87 & 13 45 42.81 & 27 22 19.5 & 1.183  &  -25.69  & QSO & 18.98 & -0.72 &   0.36 &   2.42 &   2.39 &  2.57 \\
 88 & 13 45 46.05 & 28 09 17.0 & 2.225  &  -26.62  & QSO & 19.79 & -0.50 &   0.20 &   0.89 &   1.43 &  1.33 \\
 89 & 13 45 54.54 & 27 41 01.4 & 1.038  &  -25.25  & QSO & 19.11 & -0.71 &   0.46 &   1.81 &   2.13 &  1.64 \\
 90 & 13 46 06.68 & 27 11 22.2 & 1.806  &  -25.67  & QSO & 20.13 & -1.05 &   0.52 &   2.05 &   1.86 &	  * \\
 91 & 13 46 14.57 & 28 13 52.5 & 0.659  &  -24.75  & QSO & 18.43 & -0.39 &   0.21 &   1.62 &   2.37 &  4.81 \\
 92 & 13 46 33.56 & 27 05 58.3 & 2.340  &  -27.23  & QSO & 19.32 & -0.21 &   0.16 &   0.54 &   1.60 &  1.07 \\
 93 & 13 46 38.35 & 27 57 41.2 & 2.439  &  -27.61  & QSO & 19.09 &  0.03 &   0.15 &   2.07 &   1.64 &  1.73 \\
 94 & 13 46 42.71 & 27 16 12.9 & 1.612  &  -25.78  & QSO & 19.67 & -0.49 &   0.28 &   1.18 &   2.19 &  2.65 \\
 95 & 13 46 44.32 & 28 01 30.0 & 1.127  &  -25.68  & QSO & 18.86 & -0.69 &   0.45 &   1.30 &   3.84 &  6.32 \\
 96 & 13 47 02.80 & 26 59 35.1 & 1.504  &  -25.63  & QSO & 19.66 & -0.50 &   0.26 &   1.47 &   2.62 &  3.06 \\
 97 & 13 47 03.74 & 27 09 24.8 & 1.932  &  -26.70  & QSO & 19.32 & -0.69 &   0.39 &   1.47 &   3.56 &  1.26 \\
 98 & 13 47 05.51 & 28 18 05.0 & 0.255  &  -21.20  & NELG& 19.83 & -0.36 &   1.17 &   0.57 &   0.92 &  1.52 \\
 99 & 13 47 10.23 & 28 03 54.6 & 0.992  &  -25.26  & QSO & 19.00 & -0.45 &   0.29 &   1.05 &   1.85 &  1.36 \\
100 & 13 47 16.51 & 27 14 20.1 & 2.530  &  -27.21  & QSO & 19.65 & -0.41 &   0.34 &   3.21 &   1.65 &  1.52 \\
101 & 13 47 26.33 & 27 04 33.1 & 2.212  &  -27.66  & QSO & 18.74 & -0.24 &   0.31 &   3.07 &   1.75 &  3.07 \\
102 & 13 47 30.86 & 27 13 58.5 & 2.118  &  -26.68  & QSO & 19.62 & -0.68 &   0.42 &   1.08 &   1.64 &  2.80 \\
103 & 13 48 05.16 & 27 47 13.1 & 1.430  &  -26.14  & QSO & 19.02 & -0.53 &   0.10 &   1.87 &   1.99 &  2.12 \\
104 & 13 48 25.07 & 27 06 16.4 & 2.600  &  -29.33  & QSO & 17.66 & -0.20 &  -0.06 &   1.26 &   2.58 &  8.82 \\
\bottomrule
\end{tabular}	
\addtocounter{table}{-1}	 
\caption{
continued}
}
\end{table*}

%

%
\begin{acknowledgements}
%
This research is based on observations made with the
2.2m telescope of the German-Spanish Astronomical Centre, Calar Alto, Spain.
This research has made use of the NASA/IPAC Extragalactic
Database (NED) which is operated by the Jet
Propulsion Laboratory, California Institute of Technology, under
contract with the National Aeronautics and Space Administration.

\end{acknowledgements}



\begin{thebibliography}{}

   \bibitem[1993]{Ant93} Antonucci, R. 1992, ARA\&A 31, 473

   \bibitem[2000]{Bal00} Ball, M. 2000, Ph.D. Thesis, University Jena

   \bibitem[2000]{Bec00} Becker, R. H., White R. L., Gregg M. D., et al. 
   2000, ApJ 538, 72

   \bibitem[1997]{Bec97} Becker, R. H., Gregg, M. D., Hook, I. M., et al.
   1997, ApJ 479, L93

   \bibitem[1998]{Ber98} Bershady, M. A., Trevese, D., \& Kron, R. G. 
   1998, ApJ 496, 103 

   \bibitem[1990]{Boy90} Boyle, B. J., Fong, R., Shanks, T., et al. 1990, MNRAS    243, 1

   \bibitem[2001]{Bru01} Brunzendorf, J., \& Meusinger, H. 2001, A\&A 373, 38

   \bibitem[2002]{Bru02} Brunzendorf, J., \& Meusinger, H. 2002, A\&A, in press

   \bibitem[1998]{Con98} Conti, A., Kennefick, J. D., Martini, P., et al. 1998,
   AJ 117, 645

   \bibitem[1990]{Cra90} Crampton, D., Cowley, A. P., \& Hartwick, F. D. A.
   1990, AJ 100, 47


   \bibitem[2001]{Cro01} Croom, S. M., Smith R. J., Boyle, B. J., et al. 2001,
   MNRAS, 322, L29


   \bibitem[2000]{Fra00} Francis, P. J., Whiting, M. T., \& Webster, R. L.
    2000, Publ. Astron. Soc. Aust., 53, 56

   \bibitem[1998]{Gol98} Goldschmidt, P. \& Miller, L. 1998, MNRAS 293, 107
   
   \bibitem[2002]{Gre02} Gregg, M. D., Lacy, M., White, R., et al. 2002,
   ApJ 564, 133

   \bibitem[2002]{Hal02} Hall, P. B., Anderson, S. F., Strauss, M. A., 
   et al., ApJS, in press (astro-ph/0203252v1)

   \bibitem[1990]{Har90} Hartwick, F. D. A., \& Schade, D., 1990,
   ARA\&A 28, 437

   \bibitem[1998]{Has98} Hasinger, G., Burg, R., Giacconi, R., et al. 1998,
   A\&A 329, 482

   \bibitem[1983]{Haw83} Hawkins. M. R. S. 1983, MNRAS 202, 571
 
   \bibitem[1995]{Hew95} Hewett, P. C., Foltz, C. B., \& Chaffee, F. H. 1995, AJ 109, 14   98

   \bibitem[1994]{Hoo94} Hook, I. M., McMahon, R. G., Boyle, B. J., et al.
   1994, MNRAS 268, 305


   \bibitem[1992]{LaF92} La\,Franca, F., Cristiani, S., \&
   Barbieri, C. 1992, AJ 103, 1062

   \bibitem[1981]{Kro81} Kron, R. G., \& Chiu, L.-T. 1981, PASP 93, 397 
      
   \bibitem[1997]{LaF97} La\,Franca, F. \& Cristiani, S. 1997, AJ 113, 11517

   \bibitem[1997]{Lam97} Lamer, G., Brunner, H., \& Staubert, R. 1997, A\&A 327, 467

   \bibitem[2001]{Mai01} Maiolino, R., Salvati, M., Marconi, A., et al.
   2001, A\$A 375, 25

   \bibitem[1991]{Maj91} Majewski, S. R., Munn, J. A., Kron, R. G., et al. 
   1991, in ASP Conf. Ser. 21, The Space Distribution of Quasars, 
   ed. D. Crampton, 55   

   \bibitem[2001]{Men01} Menou, K., Vanden Berk, D. E., Iveci\'c, \v{Z}. et al.
    ApJ 561, 645 
   
   \bibitem[1995]{Meu95} Meusinger, H., Klose, S., Ziener, R., et al. 
   1995, in ASP Conf. Ser. 84, The Future Utilisation of Schmidt Telescopes,
   Proceedings of IAU Colloquium 148, 
   ed. J. M. Chapman et al., 486  
    
   \bibitem[1997]{Meu97} Meusinger, H., Brunzendorf, J., Scholz, R.-D., et al.
   1997, in Treasure Hunting in Astronomical Plate Archives, ed. P. Kroll et al.
   (Thun \& Frankfurt: Harri Deutsch), 122 

   \bibitem[2001]{Meu01} Meusinger, H., \& Brunzendorf, J. 2001, A\&A 374, 878

   \bibitem[2002]{Meu02b} Meusinger, H., \& Brunzendorf, J. 2002, A\&A,
   in press

   \bibitem[2002]{Meu02a} Meusinger, H., Brunzendorf, J., Scholz, R.-D., et al.
   2002, in ASP Conf. Ser., AGN Surveys, Proceedings of IAU Colloquium 184,
   ed. R.F. Green, E.Ye. Khachikian, \& D.B. Sanders, in press

   \bibitem[2001]{Ris01} Risaliti, G., Marconi, A., Maiolino, R., et al.
   2001, A\$A 271, 37

   \bibitem[1988]{San88} Sanders, D., Soifer, B. T., Elias, J. H., et al. 
   1988, ApJ 325, 74

   \bibitem[2002]{Sch02} Schneider, D. P., Richards, G. T., Fan, X., et al.
   2002, AJ, 123, 567
   
   \bibitem[1997]{Sch97} Scholz, R.-D., Meusinger, H., \& Irwin, M.  1997,
   A\&A 325, 457 (Paper\,1)

   \bibitem[2001]{Sha01} Sharp, R. G., McMahon, R. G., Irwin, M. , et al.
   2001, MNRAS 326, 45

   \bibitem[1997]{Skr97} Skrutskie, M. F., Schneider, S. E., Stiening, R.,
   et al. 1997, in The Impact of Large Scale Near-IR Sky Surveys, ed.
   F. Garzon et al. (Dordrecht: Kluwer Academic Publishing Company), 25

   \bibitem[1995]{Ver95} V\'eron, P., \& Hawkins, M. R. S. 1995, A\&A 296, 665 

   \bibitem[2001]{Ver01}V\'eron-Cetty M. P., \& V\'eron P. 2001,
   Quasars and Active Galactic Nuclei (10th Ed.), A\&A 374, 92


   \bibitem[1995]{Web95} Webster, R. L., Francis, P. J., Peterson, B. A., et al. 
   1996, Nature 375, 469

   \bibitem[1991]{Wey91} Weymann, R. J., Morris, S. L., Foltz, C. R., et al.
   1991, ApJ 373, 23

   \bibitem[2000]{Whi00} White, R. L., Becker, R. H., Gregg, M. D., et al. 2000, 
   ApJS 126, 133

   \bibitem[1998]{Wis98} Wisotzki, L. 1998, Astron. Nachr. 319, 5

   \bibitem[2000]{Wis00} Wisotzki, L., Christlieb, N., Bade, N., et al. 2000, 
   A\&A, 358, 77

\end{thebibliography}
\end{document}